\documentclass[twocolumn,pra,superscriptaddress]{revtex4}
\usepackage[utf8]{inputenc}
\usepackage{amsmath}
\usepackage{amssymb}
\usepackage{bm}
\usepackage{graphicx}
\usepackage{color}

\begin{document}

\title{Critical current for an insulating regime of an underdamped current-biased topological Josephson junction}
\author{Aleksandr E. Svetogorov}
\author{Daniel Loss}
\author{Jelena Klinovaja}
\affiliation{University of Basel, Department of Physics, Klingelbergstrasse, 4056 Basel, Switzerland}

\date{\today}
\begin{abstract}
We study analytically an underdamped current-biased topological Josephson junction. First, we consider a simplified model at zero temperature, where the parity of the non-local fermionic state formed by Majorana bound states (MBSs) localized on the junction is fixed, and show that a transition from insulating to conducting state in this case is governed by single-quasiparticle tunneling rather than by Cooper pair tunneling in contrast to a non-topological Josephson junction. This results in a significantly lower critical current for the transition from insulating to conducting state. We propose that, if the length of the system is finite, the transition from insulating to conducting state occurs at exponentially higher bias current due to hybridization of the states with different parities as a result of the overlap of MBSs localized on the junction and at the edges of the topological nanowire forming the junction. Finally, we discuss how the appearance of MBSs can be established experimentally by measuring the critical current for an insulating regime at different values of the applied magnetic field.
\end{abstract}
\maketitle

\section{Introduction}
Topological superconductors have recently received much attention in the condensed matter community as a new exotic form of quantum matter~\cite{Wilczek2009,Alicea2012,Beenakker2013} and, moreover, as prospective candidates for quantum computation schemes due to the non-Abelian nature of Majorana fermions, which are formed at edges of such systems~\cite{Kitaev2001,Nayak2008,Alicea2011,Hoffman2016,Plugge2017}. 
However, even the direct observation of these states presents a challenging problem, which is still under active investigation~\cite{Stanescu2019,Reeg2018,Frolov2020,Sarma2018,Moore2018,Prada2018,Prada2019}. 
In this paper, we discuss effects that can indicate the existence of MBSs in topological Josephson junctions and supplement often ambiguous zero-bias peak signatures. 

\begin{figure}
\includegraphics[width=2.7cm]{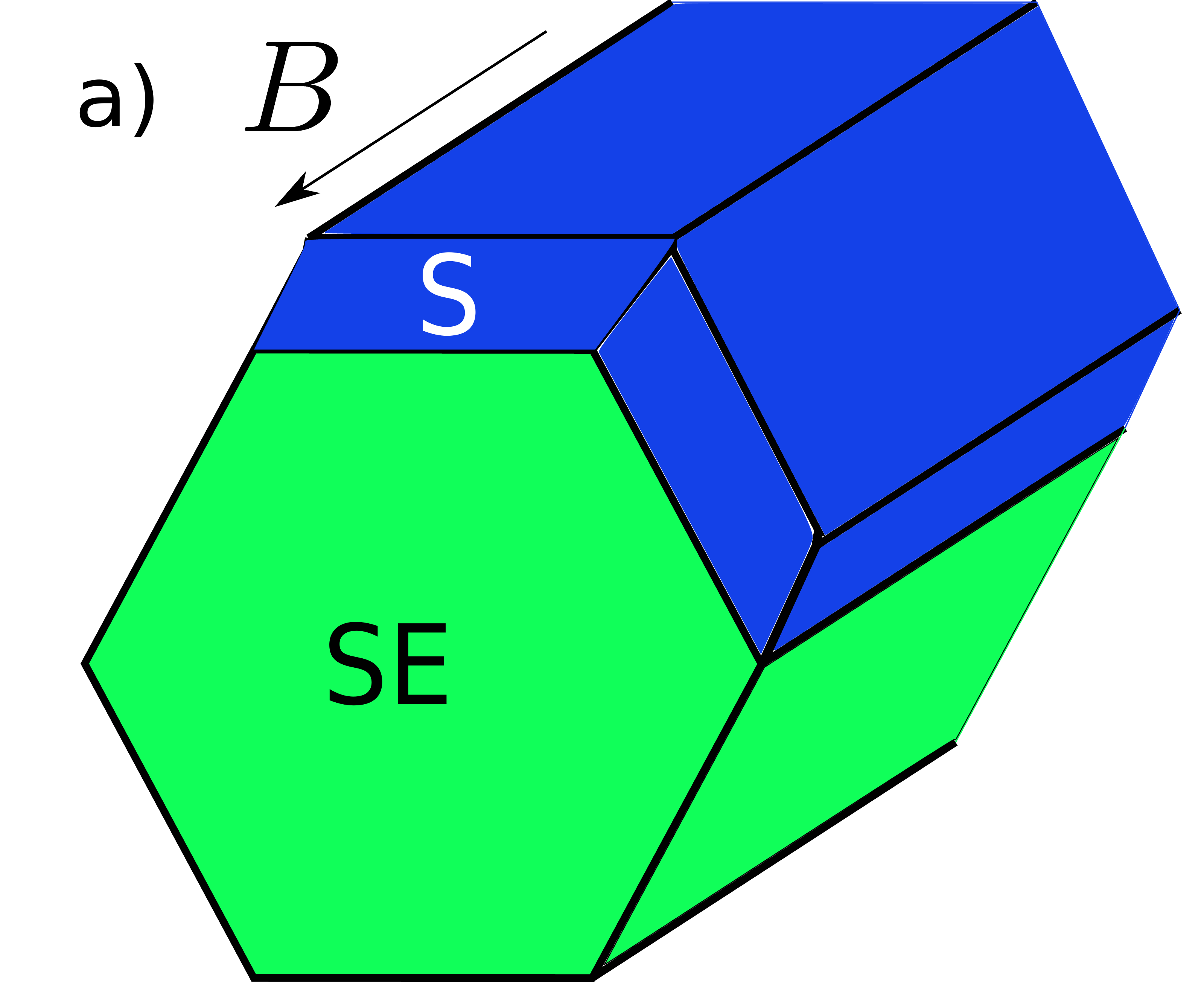}
\includegraphics[width=2.7cm]{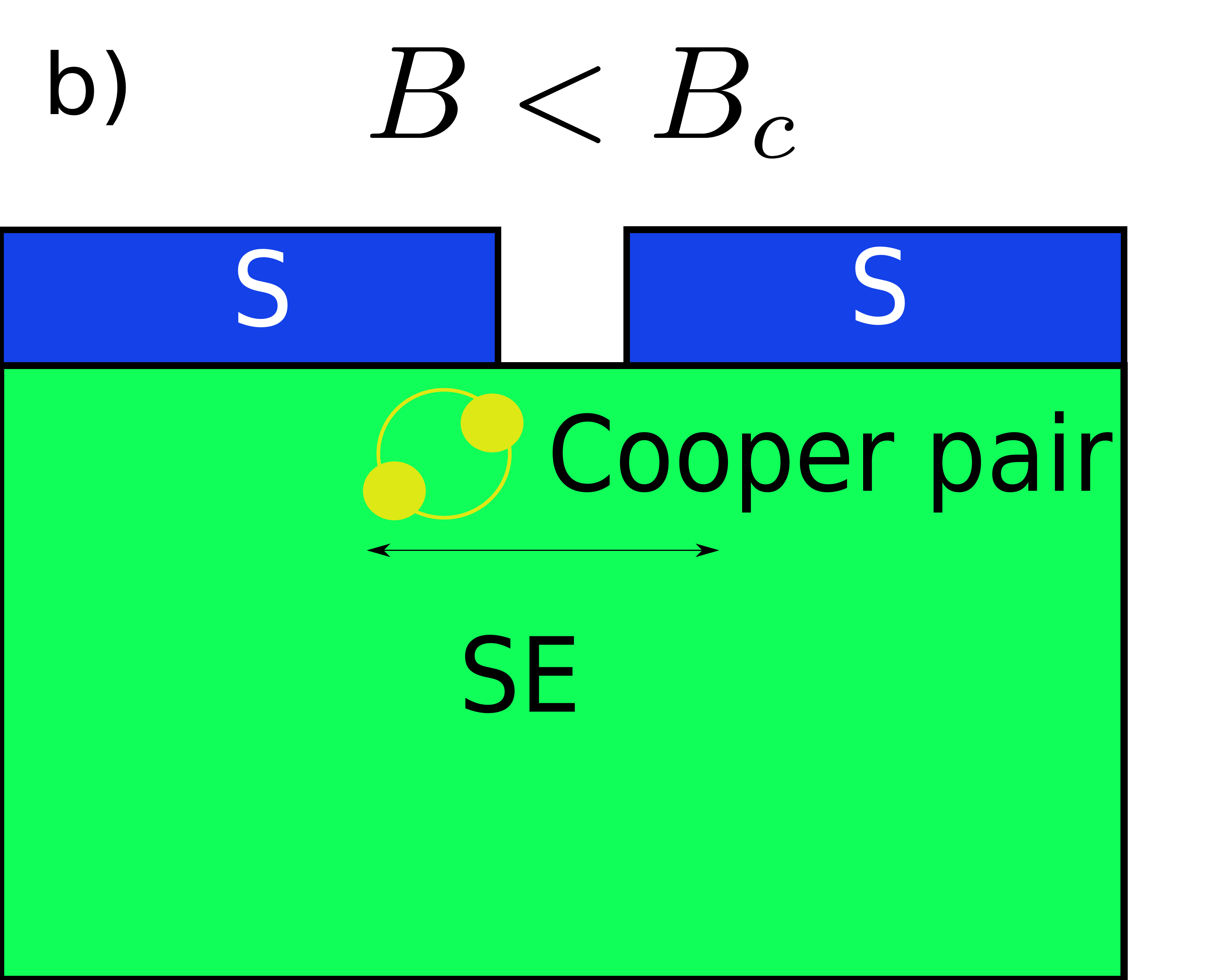}\quad
\includegraphics[width=2.7cm]{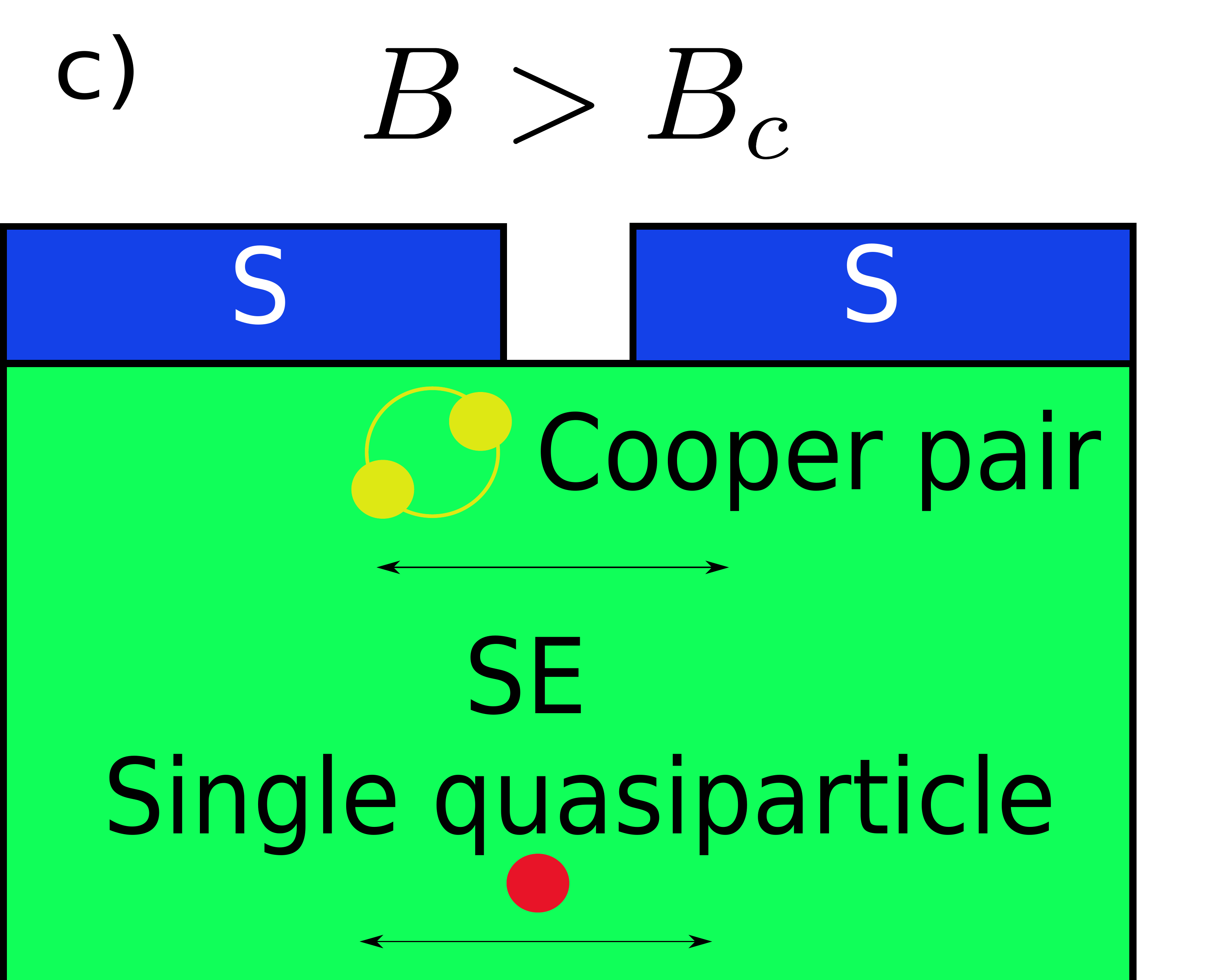}
\caption{\label{fig:Junction}
Schematic representation of the system: a) cross section of a semiconducting nanowire (SE, green) with layers of superconductor (S, blue) on two facets and magnetic field $B$ applied along the nanowire, b) a Josephson junction in the non-topological state ($B<B_c$, only Cooper pairs can tunnel), and c) a Josephson junction in the topological regime ($B>B_c$, with competition between Cooper-pair and single-quasiparticle tunneling). }
\end{figure}

There are several platforms to fabricate a topological Josephson junction: topological insulators~\cite{Ioselevich2011,Kurter2015,Wiedenmann2016}, semiconducting nanowires~\cite{Oreg2010,Sau2010,San-Jose2013,Albrecht2016,Nesterov2016,deMoor2018}, quantum dots~\cite{Schrade2017}, quantum spin-Hall insulators~\cite{Dolcini2015} or even more exotic ones like carbon nanotubes~\cite{Klinovaja2012,Sau2013,Marganska2018}. In this paper we restrict ourselves to a model of a  semiconducting single-channel nanowire with strong spin-orbit interaction in the presence of a strong  magnetic field applied along the nanowire axis, which results in two split  subbands in the nanowire~\cite{Oreg2010,Alicea2010}. The nanowire is assumed to be proximity-coupled to a conventional $s$-wave superconductor, which effectively induces $p$-wave pairing. Typically, in experimental setups, the semiconducting nanowire has a hexagonal cross section, the $s$-wave superconductor is a thin layer covering few facets of the nanowire~\cite{Krogstrup2015,Albrecht2016,Deng2016,deMoor2018}. As a result, the topological state exists at magnetic fields larger than the critical value $B>B_c=\frac{2}{g\mu_B}\sqrt{\Delta^2+\mu^2}$, determined by the superconducting pairing term $\Delta$ induced by the proximity effect  and by the chemical potential $\mu$, where the constants $g$ and $\mu_B$ are Landé $g$ factor and Bohr magneton, respectively.
The Josephson junction can be realized if a part of the nanowire is not covered by a superconducting layer (Fig.~\ref{fig:Junction}) or if there is a thin insulating segment being inserted in the superconducting layer. In the first realization, the effective Josephson junction is dominated by single-quasiparticle tunneling via the MBSs on the sides of the junction if the junction has low transparency~\cite{Kwon2003,Fu2009,Pekker2018,Aguado2020_1}. For high transparency junctions, the conventional Cooper-pair tunneling dominates. In the second realization, there is also an additional contribution to Cooper-pair tunneling due to possibility of tunneling through an insulating strip. Therefore, it may be possible to have  Cooper-pair tunneling dominating even for not very transparent Josephson junctions. We start with the system whose length is large enough to neglect the effects of the MBSs at the outer edges of the nanowire on the Josephson junction (finite-size effects are discussed in Sec.~\ref{sec:finite}). Then the Hamiltonian of the system can be written as~\cite{Likharev1985,Oppen2011,Pekker2018} (we put $\hbar=1$ throughout the paper)
\begin{equation}
H=\frac{q^{2}}{2C}+{H}_M-E_{J}\cos\phi
-\frac{\left(I-I_{q}\right)\phi}{2e}+H_{q},
\label{eq:H1}
\end{equation}
where $q$ is the electric charge  on the Josephson junction of capacitance $C$, $\phi$ is the superconducting phase difference across the junction, $E_J$ is the Josephson energy of the junction. The last two terms in Eq.~(\ref{eq:H1}) account for the driving current $I$ and the dissipation through a large impedance shunting the junction, respectively. Here, $I_q$ is the current through this impedance and $H_q$ the Hamiltonian of a thermal bath, representing the dissipation in the impedance. Two MBSs on the sides of the Josephson junction are described by $H_M=\frac{1}{2}E_M\Gamma\cos\left(\phi/2\right)$: $\Gamma$ can be associated with the parity of the state formed by these MBSs ($\Gamma=\pm1$ for odd and even parity, respectively), $E_M$ is the coupling energy between the MBSs on the junction \cite{Zyuzin2013} and characterizes the single-quasiparticle tunneling through the junction. This $H_M$ represents an effective two-level system, where the levels correspond to the occupation of an effective non-local fermionic state formed by the left and right MBSs localized on the sides of the junction. As a result, each parity is associated with the occupation of this fermionic state. We consider the junction in the limit when the phase $\phi$ is well defined. Therefore, the terms corresponding to electron tunneling should dominate over the Coulomb interaction terms, i.e., $E_M\gg{E}_c=e^2/(2C)$ and $E_J\gg{E_c}$. 

In this work we study the initial part of the current-voltage dependence for an underdamped topological Josephson junction. It is known that at low currents a Josephson junction shunted by large impedance $\mathrm{Re}\,Z>Z_Q=2\pi/(2e)^2$ (underdamped junction) is in a zero-current Coulomb blockade state (effectively insulating) due to quantum phase fluctuations~\cite{Likharev1985,Ioffe2007,Zazunov2008}. The voltage $V$ depends linearly on the current $I$ as the current flows through the external impedance $Z$; this regime holds up to some critical current $I_c$, which depends on the lowest band dispersion of a junction. In a topological junction this lowest band dispersion should be significantly different from a non-topological case, which should be seen in the value of this critical current $I_c$. The idea of an equilibrium measurement seems to be especially promising in comparison to dynamical detection schemes, as the evidence of $4\pi$ effects in non-topological junctions has been shown recently in dynamical experiments, i.e. missing Shapiro steps~\cite{Shabani2020}, which is supposed to be the result of Landau-Zener transitions. While in equilibrium measurements there are no Landau-Zener transitions between Andreev bound states, $4\pi$ periodicity can still be seen as a special property of a topological junction. We do not consider the opposite limit of overdamped Josephson junction in this work, as strong dissipation results in phase localization and, therefore, no effectively insulating regime for a current-biased junction emerges~\cite{Zazunov2008}. We consider the temperature to be sufficiently low (much lower than level spacing $\omega_0$, see Eq.~[\ref{eq:omega_M},\ref{eq:omega_J}]). In principal, thermal fluctuations should smear the voltage peak $V_c=ZI_c$~\cite{Zazunov2008}, however, the probability of thermally activated phase slips is exponentially low for such a temperature regime, therefore, we neglect the corrections due to finite temperature in this work.

The paper is organized as follows. In Sec.~\ref{sec:fixed} we introduce the simplified model with the fixed fermionic parity, which corresponds to an infinite nanowire limit. We derive the expressions for the lowest band of a topological Josephson junction in two important limits: $E_M\gg{E}_J$ and $E_M\ll{E}_J$, and calculate the critical current for an insulating regime of the Josephson junction. In Sec.~\ref{sec:finite}, we discuss finite-size effects. We show that the critical current in this regime is significantly larger, however, it is possible that at certain values of the applied magnetic field the critical current falls to the values characteristic for  infinite systems. We summarize our results and give an outlook in Sec.~\ref{sec:summary}. In App.~\ref{app:instanton} we discuss the instanton action and the fluctuation determinant for our problem.

\section{Fixed parity state}\label{sec:fixed}

Let us start with the simplified model of a very long nanowire, introduced in the previous section, so that we can neglect the overlap between MBSs on the junction and MBSs on the edges of the wire. At zero temperature and without quasiparticles, we can consider the fermionic parity to be fixed. Without loss of generality we can choose an odd parity state. Let us start with the case of zero bias current and no dissipation. Having fixed the parity, we can integrate out the degrees of freedom corresponding to the subgap fermion formed by the MBSs localized on the junction. The effective Hamiltonian takes the form~\cite{Pekker2018}
\begin{equation}
\hat{H}=\frac{q^{2}}{2C}-\frac{E_M}{2}\cos\frac{\phi}{2}-E_{J}\cos\phi.
\end{equation}
In analogy with a particle moving in a one-dimensional periodic potential~\cite{Likharev1985}, the first term in this Hamiltonian may be seen as  kinetic energy, while
\begin{equation}\label{eq:V}
V (\phi)=-\frac{E_M}{2}\cos\frac{\phi}{2}-E_{J}\cos\phi
\end{equation}
is the potential energy (the phase difference $\phi$ plays the role of the conjugate coordinate), which is depicted in Fig.~\ref{fig:potential}.

In a non-topological junction with $E_M=0$, the spectrum consists of energy bands due to coherent $2\pi$ phase slips~\cite{Likharev1985}. In the topological junction, the picture is slightly different. In the regime where
 single-quasiparticle tunneling dominates over Cooper-pair tunneling ($E_M\gg{E}_J$), the band structure is determined by $4\pi$ phase slips. In the opposite limit ($E_M\ll{E}_J$), the band structure is either determined by $4\pi$ or $2\pi$ phase slips, depending on the interplay between $E_M$ and $\nu_{0}$, which is the tunneling amplitude between the neighboring minima~\cite{RodriguezMota2019}. The value of $\nu_{0}$ is defined below in Eq.~(\ref{eq:nu_0}).

\subsection{Lowest energy band for the topological junction}
We start with the case in which single-quasiparticle tunneling dominates, i.e., $E_{M}\gg E_{J}$. If we completely ignore the Josephson term,  the corresponding Schr\"odinger equation becomes 
\begin{equation}
\frac{d^{2}}{d\left(\phi/2\right)^{2}}\psi+\left(\frac{E}{E_{c}}+\frac{E_{M}}{2E_{c}}\cos\frac{\phi}{2}\right)\psi=0,
\end{equation}
which is the Mathieu equation. The wave functions  $\psi$ corresponding to that equation should be composed of Bloch wave functions:
\begin{equation}
\psi(\phi)=\sum_n\int{d}k\ C_k^{(n)}\psi_k^{(n)},\quad\psi_k^{(n)}=u_k^{(n)}(\phi)e^{ik\phi},
\end{equation}
where $u_k^{(n)}(\phi)$ is $4\pi$-periodic and $n$ corresponds to the band number. As we are looking for the lowest bands in the limit $E_M\gg{E_c}$, we can use the tight-binding approximation and present $u_k^{(n)}(\phi)$ in the Wannier form,
\begin{equation}
u_k^{(n)}(\phi)=\sum_{m=-\infty}^\infty{w}^{(n)}\left(\phi-4\pi{m}\right)e^{-i(\phi-4\pi{m})k},
\end{equation}
where $w^{(n)}(\phi)$ are the eigenfunctions of the harmonic oscillator with the frequency $\omega_0=\sqrt{E_M{E}_c}$. This gives us the bands dispersion
\begin{equation}\label{eq:bs1}
E^{(n)}(k)=\omega_{0}\left(n+\frac{1}{2}\right)+2\left(-1\right)^{n+1}\nu^{(n)}_{4\pi}\cos(4\pi{k})
\end{equation}
with exponentially small amplitudes (which correspond to coherent tunneling between the $n$-th states in two neighboring minima of the potential~\cite{Coleman1977})
\begin{equation}
\nu^{(n)}_{4\pi}=\sqrt{\frac{2}{\pi}}E_{c}\left(\frac{E_{M}}{E_{c}}\right)^{n/2+3/4}\frac{2^{4n+1}}{n!}e^{-4\sqrt{\frac{E_{M}}{E_{c}}}}.
\end{equation}
This expression is valid for the lowest bands, which are close to the energy of the harmonic oscillator with frequency $\omega_0$: $n\ll{E}_M/\omega_0=\sqrt{E_M/E_c}$.

\begin{figure}[h]
\includegraphics[width=7cm]{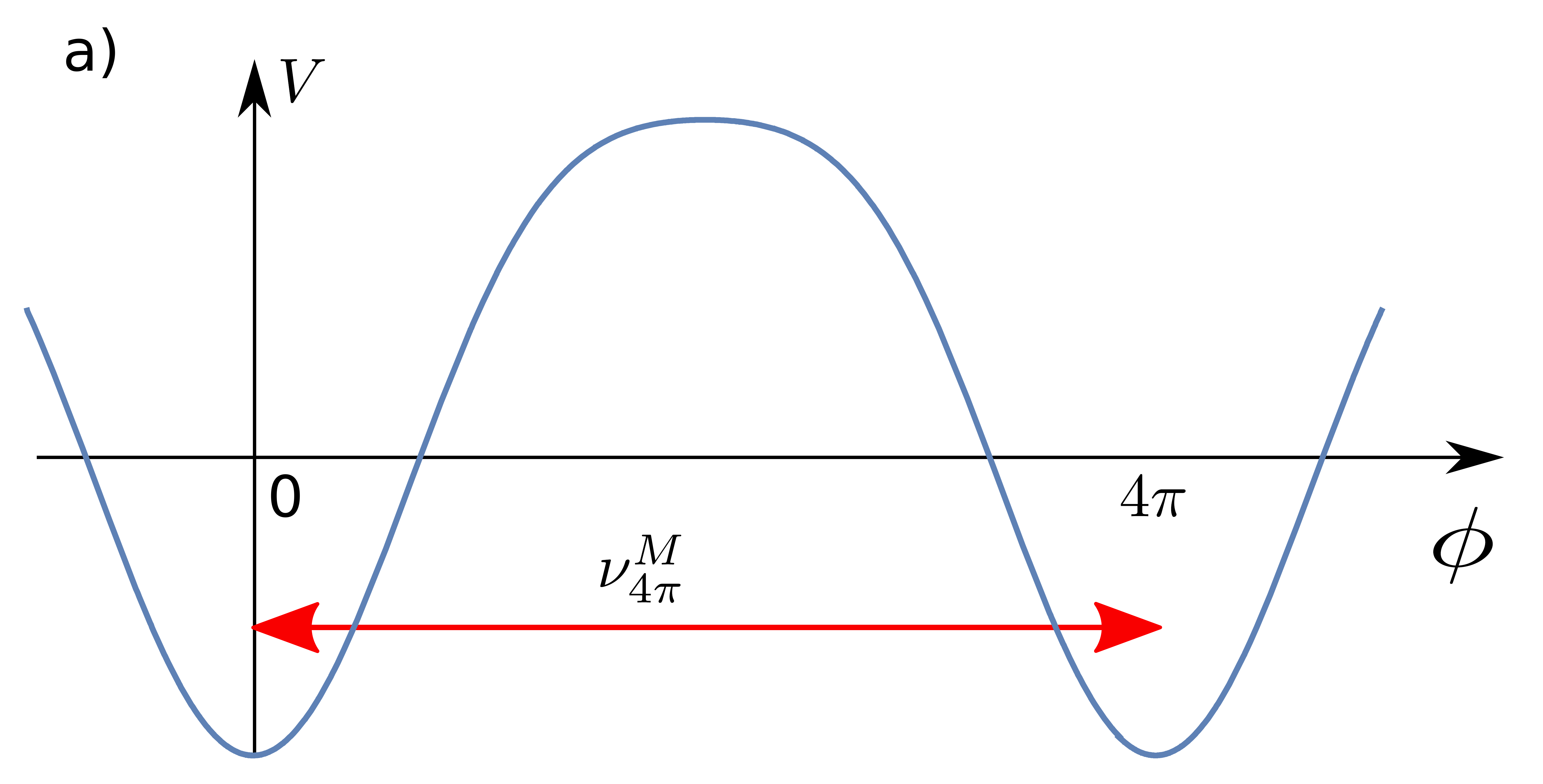}\\
\includegraphics[width=7cm]{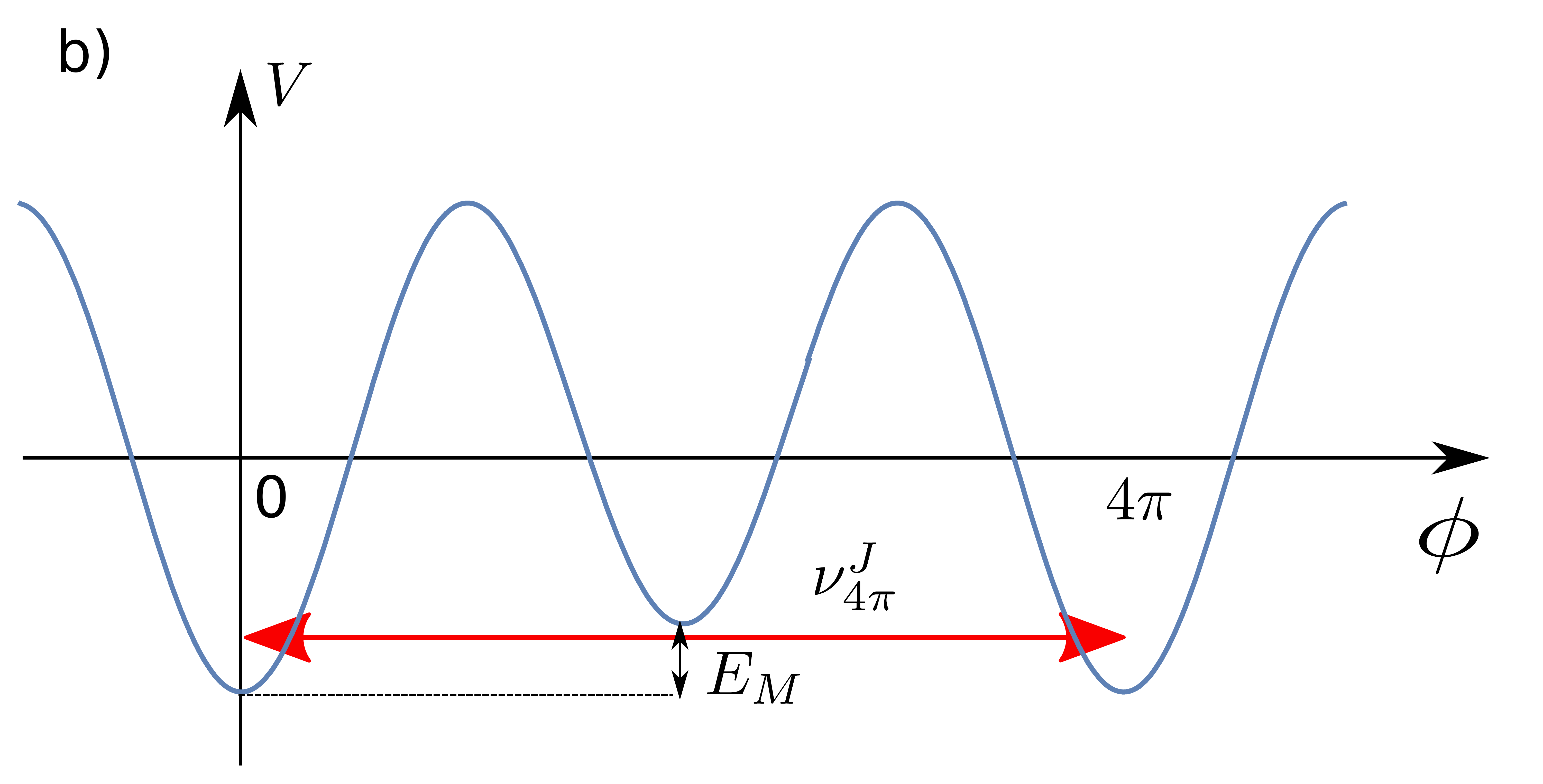}
\caption{The effective potential energy $V (\phi)$ as a function of
the phase difference $\phi$, see Eq. (\ref{eq:V}). We schematically indicate the $4\pi$ tunneling between minima of an effective potential in the two limits: a) $E_M\gg{E}_J$ and b) $E_M\ll{E}_J$. In the latter limit the potential also exhibits a set of local minima, shifted from the absolute minima by $E_M$. }
\label{fig:potential}
\end{figure}

Including the Josephson term into our consideration perturbatively will modify the harmonic frequency to \begin{multline}\label{eq:omega_M}
\omega_0=\sqrt{\left(E_M+8E_J\right)E_c}\\
=\sqrt{E_ME_c}\left[1+4\frac{E_J}{E_M}+O\left(\frac{E_J}{E_M}\right)^2\right]
\end{multline}
 as well as the exponent, determined by an instanton action (see Appendix~\ref{app:instanton}), connecting neighboring minima of the potential (see Fig.~\ref{fig:potential}). 
We neglect the correction to the pre-exponential term in the amplitude. The instanton action is given by
\begin{multline}
S^M_{4\pi}=\sqrt{\frac{E_{M}}{8E_{c}}}\intop_{0}^{4\pi}\sqrt{1-\cos\frac{\phi}{2}+\frac{2E_{J}}{E_{M}}\left(1-\cos\phi\right)}d\phi \\
=4\sqrt{\frac{E_{M}}{E_{c}}}+\frac{16}{3}\frac{E_{J}}{\sqrt{E_{M}E_{c}}}+O\left(\frac{E_{J}^2}{E_{M}^{3/2}E_{c}^{1/2}}\right).
\end{multline}
As a result, including these modifications in Eq.~(\ref{eq:bs1}), we get the lowest energy band dispersion
\begin{equation}
E^{(0)}(k)=\frac{1}{2}\omega_{0}-2\nu_{4\pi}^M\cos(4\pi{k}),
\end{equation}
with the amplitude 
\begin{equation}
\nu_{4\pi}^M=2\sqrt{\frac{2}{\pi}}E_{c}\left(\frac{E_{M}}{E_{c}}\right)^{3/4}e^{-S^M_{4\pi}}.
\end{equation}

Next, we study the case  $E_{M}\ll E_{J}$. Here, we consider the limit of $E_M\gg\nu_0$, which corresponds to the suppression of $2\pi$ phase slips, where we introduce~\cite{Likharev1985,Matveev2002}
\begin{equation}\label{eq:nu_0}
\nu_0=4\sqrt{\frac{2}{\pi}}2^{1/4}E_c\left(\frac{E_J}{E_c}\right)^{3/4}e^{-S_{2\pi}}, 
\end{equation} 
which is the $2\pi$ tunneling amplitude in case of $E_M=0$ (which corresponds to a non-topological junction), where
\begin{equation}\label{eq:S2pi}
S_{2\pi}=\sqrt{8E_J/E_c}
\end{equation}
is the instanton action for this tunneling process.
We assume that this limit is realistic  as the phase-slip amplitude is exponentially small in the chosen range of parameters ($E_J\gg{E}_c$). Therefore, the band structure is again determined by $4\pi$ phase slips. Following the same approach as in the opposite limit, we derive
\begin{equation}
E^{(0)}(k)=\frac{1}{2}\omega_{0}-2\nu_{4\pi}^J\cos\left(4\pi k\right).
\end{equation}
Here, the harmonic frequency is  given by
\begin{multline}\label{eq:omega_J}
\omega_0=\sqrt{\left(E_M+8E_J\right)E_c}  \\ 
=\sqrt{8E_JE_c}\left[1+\frac{E_M}{16E_J}+O\left(\frac{E_M}{E_J}\right)^2\right],
\end{multline}
while the tunneling amplitude is determined again by an instanton action,
\begin{equation}\label{eq:nu_4pi}
\nu_{4\pi}^J=\sqrt{\frac{S_{4\pi}^J}{2\pi}}{\cal{N}}e^{-S_{4\pi}^J}=\sqrt{\frac{S_{4\pi}^J}{S_{2\pi}}}e^{-S_{4\pi}^J+S_{2\pi}}\nu_0.
\end{equation}
Here, $\cal{N}$ is determined by the reduced determinant (with excluded zero mode) of an operator that corresponds to the second variation of the imaginary-time action (see Appendix~\ref{app:instanton} and~\cite{Vainshtein1982,Coleman1977}), therefore, $\cal{N}$ can be considered to be the same as for the case of a non-topological junction [with the relative correction  $O\left(E_M/E_J\right)$].
The instanton action for the $4\pi$ phase slip in this limit takes the form
\begin{multline}
S^J_{4\pi}=\frac{1}{2}\sqrt{\frac{E_{J}}{E_{c}}}\intop_{0}^{4\pi}\sqrt{1-\cos\phi-\frac{E_{M}}{2E_{J}}\left(\cos\frac{\phi}{2}-1\right)}\,d\phi \\
=2\sqrt{8\frac{E_{J}}{E_{c}}}+\frac{E_{M}}{\sqrt{8E_JE_c}}\left[1+5\ln2-\ln\frac{E_{M}}{E_{J}}\right] \\
+O\left(\frac{E_{M}^{2}}{E_{J}^{3/2}\sqrt{E_c}}\right).
\end{multline}

\subsection{Critical current for the insulating state of an underdamped topological junction}

\begin{figure}[t]
\includegraphics[width=6cm]{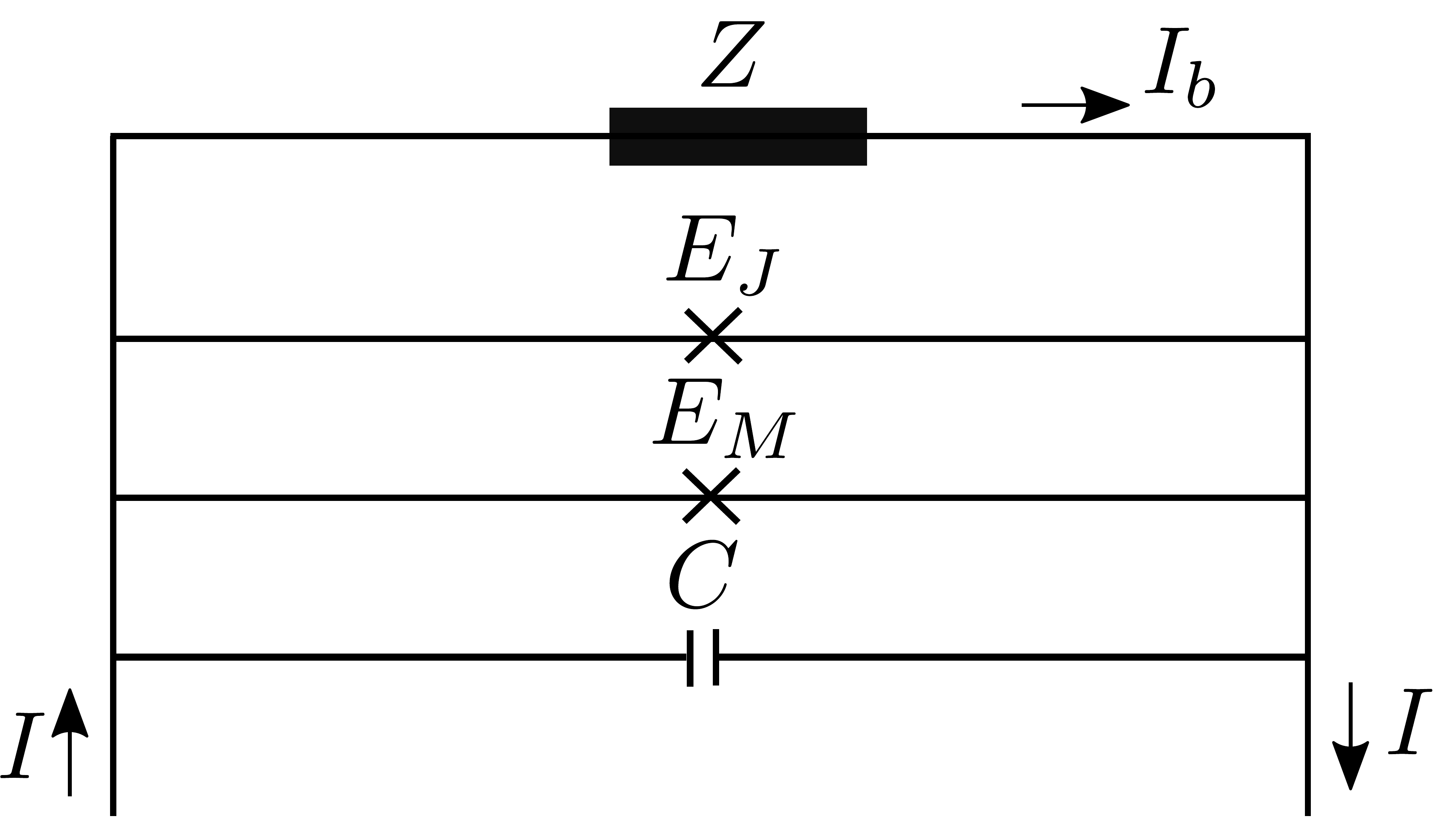}
\caption{\label{fig:circuit}
Schematic of the equivalent electric circuit for an underdamped topological Josephson junction. The applied current $I$ is divided between the shunting impedance $Z$ (current $I_b$) and the topological junction, effectively represented by the capacitance $C$, Cooper-pair tunneling element $E_J$,  and single-quasiparticle tunneling element $E_M$.}
\end{figure}

In this subsection, we study the insulating regime of an underdamped topological junction. Therefore, we include dissipation through a large impedance $Z$ into our consideration and allow for a small current $I$ through the system (see Fig.~\ref{fig:circuit}). To ensure weak dissipation, we require an underdamped junction regime: $\mathrm{Re}\,Z>Z_Q$, where $Z_Q=1/(4e^2)$ is the resistance quantum.  Using the analogy of a particle moving in a one-dimensional potential, we can write the semiclassical equations of motion~\cite{Likharev1985,Ioffe2007}: 
\begin{align}
\frac{d\phi}{dt}={}&{}\frac{dE^{(0)}}{dk},\\
\frac{dk}{dt}={}&{}\frac{I}{2e}-\frac{Z_Q}{Z}\frac{d\phi}{dt}.
\end{align}
Then, up to a critical current $I_c=2e\,\mathrm{max}\left(\frac{dE^{(0)}}{dk}\right)\frac{Z_Q}{Z}$, the current $I$ flows through the external impedance $Z$ as there is a stationary solution with constant $k$: 
\begin{equation}
\frac{d\phi}{dt}=\frac{I}{2e}\frac{Z}{Z_Q},
\end{equation}
with $V=ZI$ being the voltage. It is important to note that $I_c$ is not the maximum current supported by the junction but a critical current for an insulating regime of an underdamped junction.
This stationary regime corresponds to an insulating state of the junction. At stronger driving currents, i.e., $I>I_c$, there is no longer a solution with constant $k$ and the system enters the regime of Bloch oscillations. In this regime, for the low dissipation, the motion is periodic in $k$ \cite{Ioffe2007}.  As a result, the voltage $V$ is decreasing with the increase of the driving current $I$ and the junction is no longer in the insulating state.

We can express the critical current $I_c$ in the two limits: single-quasiparticle tunneling dominating ($E_M\gg{E}_J$) vs. Cooper-pair tunneling dominating ($E_M\ll{E}_J$).  The first limit results in the critical current
\begin{equation}
I^M_{4\pi}=32e\sqrt{2\pi}E_{c}^{1/4}E_{M}^{3/4}e^{-S^M_{4\pi}}\frac{Z_Q}{Z},
\end{equation}
while in the second limit we have
\begin{equation}
I^J_{4\pi}=128e\sqrt{\pi}2^{1/4}E_c^{1/4}E_J^{3/4}e^{-S^J_{4\pi}}\frac{Z_Q}{Z}.
\end{equation}
One can see that the expressions are sufficiently different from the one for a non-topological junction~\cite{Likharev1985,Ioffe2007}
\begin{equation}\label{eq:non-top}
I_{2\pi}=32e\sqrt{2\pi}2^{1/4}E_c^{1/4}E_J^{3/4}e^{-S_{2\pi}}\frac{Z_Q}{Z},
\end{equation}
due to an exponential factor. For $E_M\gg{E}_J$, the instanton action is parametrically larger, i.e. $S^M_{4\pi}\gg{S}_{2\pi}$, while in the opposite limit $E_M\ll{E}_J$, it is at least twice as large as in the non-topological case: 
\begin{multline}
S^J_{4\pi}=2S_{2\pi}+\frac{E_{M}}{\sqrt{8E_JE_c}}\left[1+5\ln2-\ln\frac{E_{M}}{E_{J}}\right] \\
+O\left(\frac{E_{M}^{2}}{E_{J}^{3/2}\sqrt{E_c}}\right).
\end{multline}
The critical current in both topological limits is exponentially smaller compared to the non-topological case, provided that $E_J$ can be considered to be the same as in the non-topological setup. In principle, this effect should be measurable, for example, by driving the junction from the non-topological to topological state by increasing the magnetic field. However, this increase of field will also change the effective $E_J$. We expect the first limit $E_M\gg{E}_J$ to be more promising for the demonstration of the presence of MBSs in the system, as the current $I_c$ depends mostly on $E_M$. Here, $E_J$ results only in a parametrically small corrections to the critical current. In addition, $E_M$ is non-monotonic as a function of the applied magnetic field ~\cite{Dmytruk2018,Aguado2020_1,Hoffman2017}, which results in a non-monotonic dependence of $I^M_{4\pi}$ on the magnetic field. In contrast, for a non-topological junction, $E_J$ is decreasing monotonically with the magnetic field, which results in a growth of $I_{2\pi}$ due to the exponential factor. Strictly speaking, Eq.~(\ref{eq:non-top}) should result in the growth of $I_{2\pi}$ up to some value of $B$ and further decrease due to pre-exponential factor, however, at this point the assumption $E_J\gg{E}_c$ breaks down, therefore, the above formulas are no longer valid. We expect that this should allow one to distinguish experimentally the junctions that host MBSs from those which do not. In fact, when MBSs appear, $E_J$ was also reported to show a non-monotonic dependence on the magnetic field~\cite{Aguado2020_1}. Thus, the junction in such a regime can also be used for establishing the existence of MBSs in the system. 

Unfortunately, there is another restriction for experimental observation of this effect. In any realistic experimental setup one has to take into account quasiparticles that are switching the parity of the MBSs. Therefore, this effect can be measured only on the time scales sufficiently smaller than the characteristic time $\tau_q$ between quasiparticles passing the system, while the latter could be short in existing experimental setups~\cite{Rainis2012,Schmidt2012,Higgenbotham2015,Albrecht2017,Aseev2018,Budich2012}. On the other hand, there are new  encouraging estimations for these time scales based on treating quasiparticle dynamics in finite-size one-dimensional system~\cite{Karzig2020}. Moreover, finite-size effects may change the picture dramatically; we address them in the next section.

\section{Parity switching due to finite size of the system}\label{sec:finite}

\begin{figure}
\includegraphics[width=8cm]{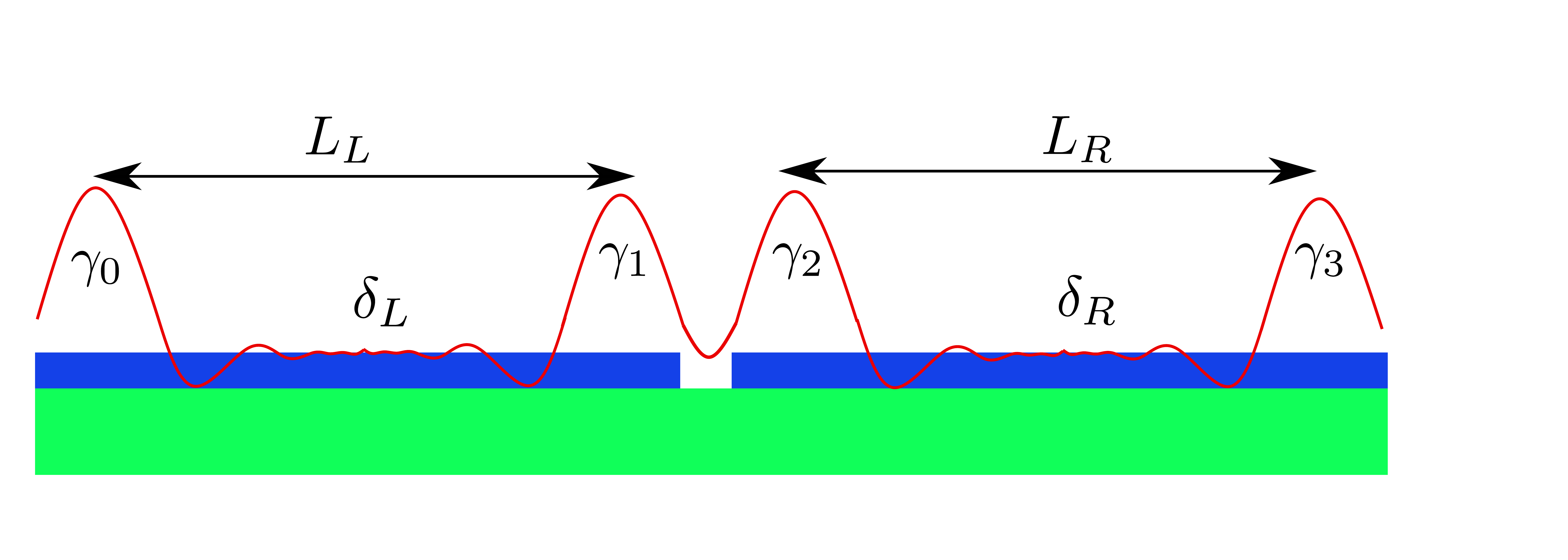}
\caption{Schematic representation of the overlap of MBSs $\gamma_{1(2)}$ on the junction and $\gamma_{0(3)}$ on the nanowire edges with associated splittings $\delta_L$ and $\delta_R$ for the left and right parts of the wire, respectively. The lengths of the corresponding parts are given by $L_L$ and $L_R$.}\label{fig:overlap}
\end{figure}

In a realistic experimental setup the whole system is finite, therefore, there is a small but finite overlap between MBSs on the junction ($\gamma_1$ and $\gamma_2$) and MBSs on the edges of the topological nanowire ($\gamma_0$ and $\gamma_3$)~\cite{Pikulin2012,Dominguez2012,Feng2018}, which results in hybridization of two states with different parities. The total parity is conserved, however, the parity of the subgap fermion formed by the MBS on the junction may change together with the parity of the non-local fermion state formed by the MBSs on the outer edges of the topological nanowire. The overlap of MBSs $\gamma_i$ is schematically depicted in Fig.~\ref{fig:overlap}. As a result, the part of the Hamiltonian $H$ [see Eq.~(\ref{eq:H1})] corresponding to the MBSs on the junction $H_M$ is modified. We can write it in the following form~\cite{Pikulin2012,San-Jose2012,Feng2018}:
\begin{equation}\label{eq:hm_splitting}
H_M=\frac{1}{2}\psi^\dagger\left(\begin{array}{cc}
E_{M}\cos\frac{\phi}{2} & \delta\\
\delta & -E_{M}\cos\frac{\phi}{2}
\end{array}\right)\psi,
\end{equation}
 where $\psi=\left(\begin{array}{c}
\psi_0\\
\psi_1
\end{array}\right)$ corresponds to the wave function of the subgap fermion state, given by $\psi_0|0\rangle+\psi_1|1\rangle$, where $|0\rangle$ and $|1\rangle$ are an even and an odd parity state, respectively  ($\left|\psi_{1}\right|^{2}+\left|\psi_{0}\right|^{2}=1$). The non-diagonal term is $\delta=\delta_L+\delta_R$, where $\delta_{L/R}$ is the coupling between the MBSs to the left/right from the junction (see Fig.~\ref{fig:overlap}).

\begin{figure}[t]
\includegraphics[width=8cm]{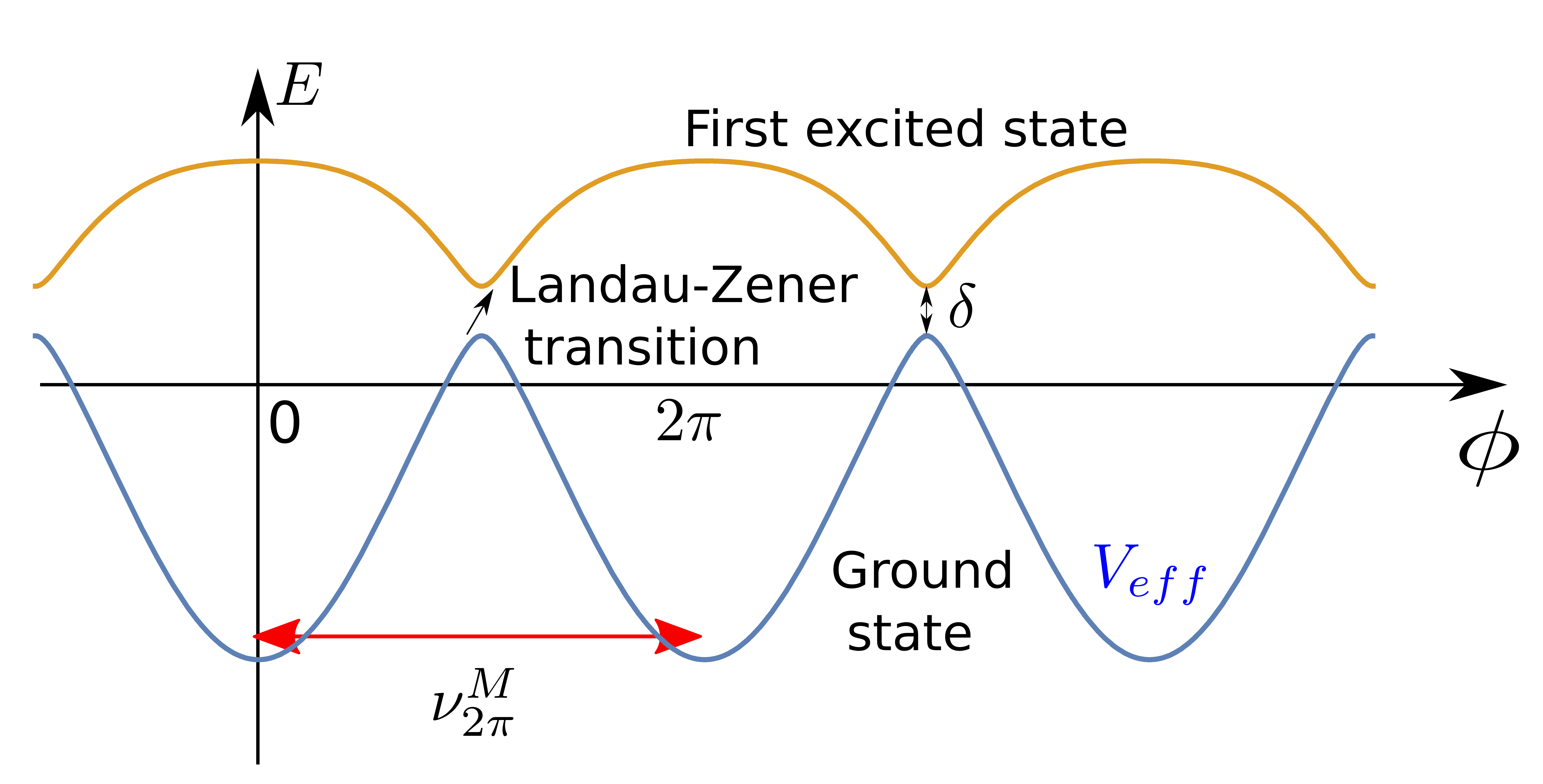}
\caption{Two lowest energy levels in a fixed phase regime (in the limit $E_M\gg{E}_J$). The ground state energy (blue curve) can be seen as an effective potential $V_{eff}$ [see Eq. (\ref{Veff})] in the adiabatic limit such that one can neglect Landau-Zener transitions. As a result, $2\pi$ phase slips are restored. }\label{fig:effective_pot}
\end{figure}

If we consider the phase to be constant, the ground state of such a system is (see Fig.~\ref{fig:effective_pot})~\cite{Pikulin2012,San-Jose2012,Feng2018}: 
\begin{equation}
E_g=-E_J\cos\phi-\frac{1}{2}\sqrt{E_M^2\cos^2\left(\phi/2\right)+\delta^2}.
\end{equation} If the total coupling energy $\delta$ is much larger than the tunneling amplitude, given by $2\pi$ phase slip, $\nu_{2\pi}$ [calculated later: Eq.~(\ref{eq:2pi_amplitude}) in two opposite limits], which gives the characteristic velocity of the phase evolution, we can consider the phase dynamics to be adiabatic in comparison to the dynamics of a two-level system, formed by MBSs on the junction, given by  Hamiltonian~(\ref{eq:hm_splitting}).  As a result, we can neglect Landau-Zener transitions at $\phi=(2n+1)\pi$, where $n$ is an integer. Then, the effective potential coincides with $E_g$ and, therefore, it is $2\pi$-periodic, which results in $2\pi$ phase slips with an amplitude higher than for $4\pi$ phase slips. The probability of the Landau-Zener transition is given by 
\begin{equation}
P_{LZ}=\exp\left(-2\pi\frac{(\delta/2)^2}{\dot{\phi}E_M/2}\right),
\end{equation}
where $\dot{\phi}=d\phi/dt$ and can be estimated as the tunneling amplitude between  neighboring minima of the effective potential $\dot{\phi}=2\pi\nu_{2\pi}$.
Therefore, the quantitative condition for this regime is 
\begin{equation}\label{eq:LZcondition}
\delta\gg\delta_c=\sqrt{2\nu_{2\pi}E_M}.
\end{equation}
That means that we can still consider $\delta$ to be sufficiently smaller than any other energy scale in the system, as the whole tunneling amplitude is exponentially small in both  limits considered, $\nu^{M/J}_{2\pi}\sim\exp\left(-S^{M/J}_{2\pi}\right)$, due to the large tunneling action. We can assume that this regime is indeed reasonable since~\cite{Rainis2013,Dmytruk2018}
\begin{equation}\delta_{L/R}\sim\frac{p_F}{m\xi_M} e^{-2L_{L/R}/\xi_M} \cos(p_FL_{L/R}),
\end{equation}
where $L_{L/R}$ is the length of the nanowire to the left/right of the junction, $\xi_M$ is the localization length of the Majorana fermions, which is of the order of hundred nanometers for typical materials like InAs, and $p_F$ is the Fermi momentum. Moreover, $p_F$ effectively grows with the applied magnetic field $B$, therefore, $\delta_{L/R}$ oscillates around zero as a function of the  magnetic field~\cite{Rainis2013,Dmytruk2018,Aguado2020_1}. As a result, experimentally it should be possible to decrease $\delta_{L/R}$ to the desirable values or even tune it to zero (that is a way to realize the limits studied in the previous section in a finite system). However, the latter assumption also relies on  $\delta_L$ and $\delta_R$ going through zero at the same values of the  magnetic field to have total splitting $\delta=\delta_L+\delta_R$ oscillating around zero. This is possible, for example, 
if the parts of the nanowire to the left and to the right of the junction are identical,
 which might be challenging to implement experimentally. Alternatively, the same effect can be achieved if, say, the left part is sufficiently long to give $\delta_L\approx 0$, while the right part is shorter with finite $\delta_R$ that can then be tuned by the magnetic field.

The effective potential takes the form (see Fig.~\ref{fig:effective_pot})
\begin{equation} \label{Veff}
V_{eff}(\phi)=-\frac{1}{2}\sqrt{E_M^2\cos^2\frac{\phi}{2}+\delta^2}-E_J\cos\phi.
\end{equation}
Then, the tunneling actions in the two opposite limits, $E_M\gg{E}_J$ and $E_M\ll{E}_J$, become
\begin{multline}
S_{2\pi}^M=\sqrt{\frac{8E_{M}}{E_{c}}}\left(\sqrt{2}-1\right)+\frac{4\sqrt{2}}{3}(2\sqrt2-1)\frac{E_{J}}{\sqrt{E_{M}E_{c}}} \\
+O\left(\frac{E_{J}^2}{E_{M}^{3/2}E_{c}^{1/2}}\right)+o\left(\delta\sqrt{\frac{E_M}{E_c}}\right)
\end{multline}
and 
\begin{multline}
S^J_{2\pi}=
\sqrt{8\frac{E_{J}}{E_{c}}}+\frac{\sqrt2 \ln2}{4}\frac{E_{M}}{\sqrt{E_JE_c}} \\
+O\left(\frac{E_{M}^{2}}{E_{J}^{3/2}\sqrt{E_c}}\right)+O\left(\frac{E_M}{\sqrt{E_JE_c}}\delta\right),
\end{multline}
 respectively.
Here, we have neglected the correction due to $\delta$, as we consider it to be small in comparison to all the energy parameters in the system except for the tunneling amplitudes.
As a result, we can calculate $\nu_{2\pi}$ for these cases
\begin{equation}\label{eq:2pi_amplitude}
\nu_{2\pi}^{M/J}=\nu_{4\pi}^{M/J}\sqrt{\frac{S_{2\pi}^{M/J}}{S_{4\pi}^{M/J}}}e^{-S^{M/J}_{2\pi}+S^{M/J}_{4\pi}},
\end{equation}
and, finally, the critical current for an insulating regime:
\begin{equation}
I^M_{2\pi}=16\left(\frac{\sqrt{2}-1}{\sqrt{2}}\right)^{1/2}e\sqrt{2\pi}E_{c}^{1/4}E_{M}^{3/4}e^{-S^M_{2\pi}}\frac{Z_Q}{Z}
\end{equation}
for $E_M\gg{E}_J$
and
\begin{equation}
I^J_{2\pi}=32e\sqrt{2\pi}2^{1/4}E_c^{1/4}E_J^{3/4}e^{-S^J_{2\pi}}\frac{Z_Q}{Z}
\end{equation}
for $E_M\ll{E}_J$.

\begin{figure}
\includegraphics[width=8cm]{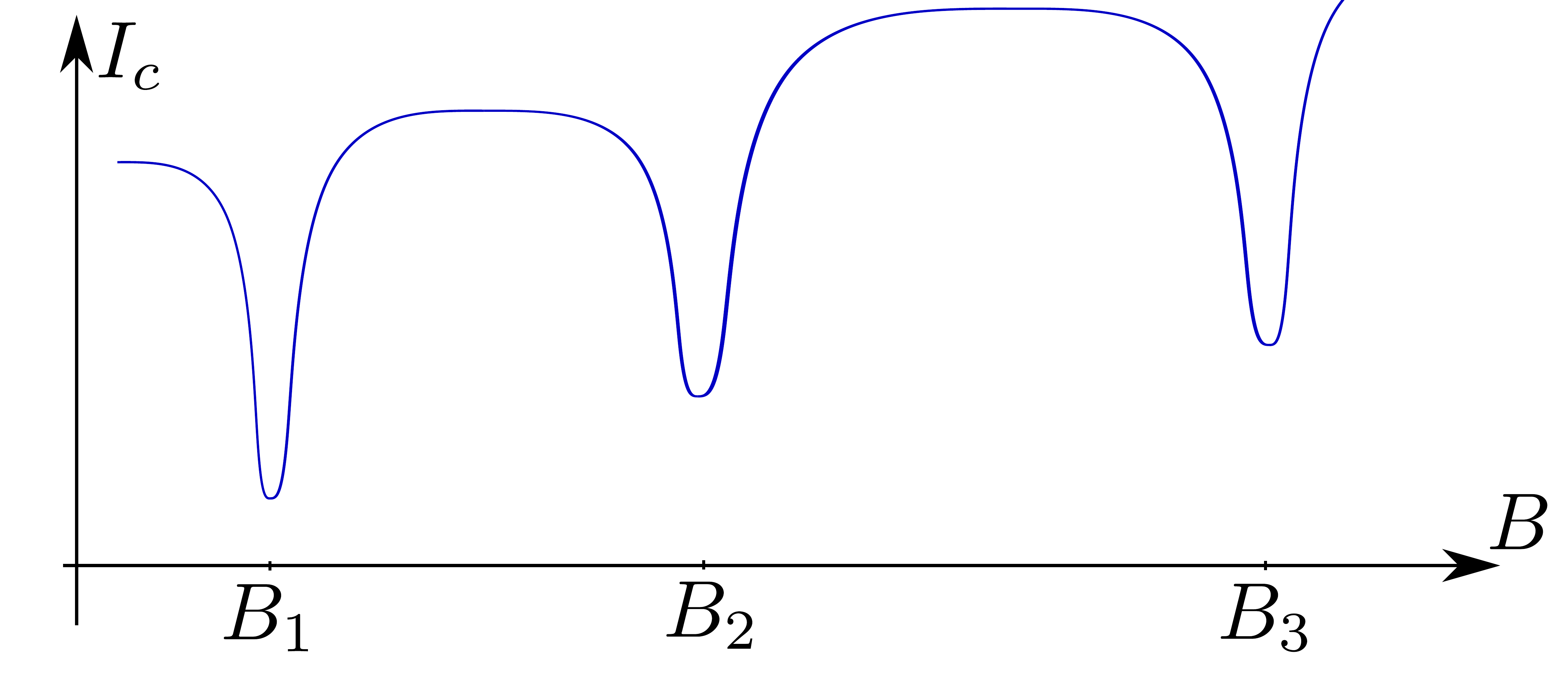}
\caption{Schematic illustration of the critical current $I_c$ as function of magnetic field $B$. At $B=B_i>B_c$, the overlap between MBSs goes to zero, $\delta=0$. As a result, the critical current $I_c$ drops exponentially. The schematic plateaus of $I_c$ correspond to $2\pi$ periodicity, while the dips correspond to (mostly) $4\pi$ periodicity of the Josephson junction. }\label{fig:ic}
\end{figure}

One can see that the critical current value in the limit $E_M\ll{E}_J$ is close to the value for the non-topological junction given in Eq.~(\ref{eq:non-top}). The reason is that the effective potential has only a parametrically weak relative modification [$O\left(E_M/E_J\right)$], while $2\pi$ phase slips are no longer suppressed. However, we note that the value of $E_J$ in topological and non-topological junctions is different and, what is more important, has a contrasting dependence on the  magnetic field. Indeed, if the system cannot support MBSs, $E_J$ decays monotonically with the magnetic field, whereas the emergence of MBSs in  magnetic fields higher than the critical value $B_c$ results in a non-monotonic dependence of $E_J$~\cite{Aguado2020_1}. In the opposite limit, the critical current for an insulating regime depends mostly on $E_M$ rather than $E_J$, which should again result in a non-monotonic dependence of $I_c$ on the magnetic field. Moreover, as $\delta_{L/R}$ oscillates around zero as a function of magnetic field, if the right and left parts of the nanowire have the same length, the total hybridization $\delta=\delta_L+\delta_R$ should also be oscillating around zero. Alternatively, again, $\delta_L$ can be made vanishingly small by increasing the length of the left part of the nanowire, while $\delta_R$ is finite and can be tuned by the magnetic field.
As a result, in some range of the magnetic field, the system should be in the limit $\delta\ll\delta_c$, which increases the probability of Landau-Zener transition to one. Therefore, the critical current should decrease dramatically due to the suppression of $2\pi$ phase slips (as shown in the previous section). This should result in a highly non-monotonic dependence of the critical current on the magnetic field, which we have schematically depicted in Fig~\ref{fig:ic}. In the proposed scheme, one should be able to distinguish the peak at voltage $V_c=ZI_c$. In the limit $\delta\ll\delta_c$, the voltage peaks may be hard to observe as the value is suppressed by the large factor in the exponent. For example, in experimentally relevant regime of proximity induced gap $\Delta=250\,{\mu{eV}}$, $E_J=0.01\Delta$, $E_M=0.02\Delta$, $E_c=0.005\Delta$ the voltage peak would be of the value of hundred nanovolts, which is on the very edge of resolution. However, for $\delta\gg\delta_c$ the factor in the exponent is smaller (and in principle can be close to the one in a non-topological junction due to restoration of $2\pi$ phase slips). For example, for the values given above the voltage peak is already of the order of ten microvolts. Therefore, the fact that at some values of the applied magnetic field the voltage peaks (as well as corresponding $I_c$) are significantly lower should be observable. 

\section{Conclusions and outlook}\label{sec:summary}
In conclusion, we have studied an underdamped topological Josephson junction. We used the effective model of a topological junction based on a semiconducting nanowire proximitized by a conventional $s$-wave superconductor. We started with deriving an expression for the lowest energy band of such a junction in the absence of a current source at zero temperature. We introduced two regimes governed either by single-quasiparticle tunneling or by Cooper-pair tunneling, which are determined by the geometry of the sample (mostly the transparency of the junction). Then we discussed the insulating regime (Coulomb blockade) of the junction, shunted by a huge impedance, which holds up to some critical bias current. We have shown that this critical current in the topological regime is sufficiently lower than in the non-topological junction with the same $E_J$ due to the possibility of single-quasiparticle tunneling and the resulting suppression of $2\pi$ phase slips. 
From an experimental point of view, a way to determine whether the junction supports MBSs or not could be to measure this critical current at different values of the magnetic field. We have argued that a non-monotonic dependence on the magnetic field  indicates the presence of MBSs. 

We continued our analysis by addressing  finite-size effects, resulting in hybridization of the states with different parities due to coupling of the MBSs on the junction with the MBSs on the outer edges of the nanowire. If the coupling energy is significantly larger than $\delta_c$, given by Eq. ~(\ref{eq:LZcondition}), the effective potential becomes $2\pi$ periodic, which results in larger tunneling amplitudes and, therefore, larger critical currents. Despite the restoration of $2\pi$ phase slips, the effective potential is still sufficiently different from the non-topological case. The main reason is that the energy scales, corresponding to the potential amplitude, have a non-trivial dependence on the applied magnetic field as mentioned above, while for non-topological junctions $E_J$ is monotonically decreasing with the field. Therefore, the same way of detecting MBSs can be used as for very long systems, where finite-size effects are negligible: the critical current for an insulating regime of the junction should show non-monotonic dependence on the magnetic field, if Majorana fermions are present.

Finally, we have also discussed a specific case where the parts of the nanowire to the right and to the left of the junction could be considered identical. Then the total hybridization energy $\delta=\delta_L+\delta_R$ should be oscillating around zero as a function of the  magnetic field. Alternatively,
$\delta_L$ can be made zero by sufficiently increasing the length of the left part of the nanowire, while the finite $\delta_R$ can be tuned by the magnetic field.
As a result, the system should move from the limit of $\delta\gg\delta_c$ to $\delta\ll\delta_c$ and back with the increase of the magnetic field. Therefore, the critical current $I_c$ for the insulating regime should have significant drops at certain values of the magnetic field (suppression of $2\pi$ phase slips). This may significantly simplify the experimental identification of MBSs in the system.

In this work we have focused on two limiting cases: $\delta\ll\delta_c$ and $\delta\gg\delta_c$, which correspond to regimes with $4\pi$ and $2\pi$ phase slips, respectively. As an outlook we plan to study 
the transition between these regimes in more detail, as the difference between these limiting cases is dramatic due to the exponentially different values of the critical current for the insulating state of an underdamped junction. 

{\it Acknowledgements.} We thank Igor Poboiko, Kirill Plekhanov, and Dmitry Miserev for fruitful discussions. This work was supported by the Swiss National Science Foundation and NCCR QSIT.
This project received funding from the European Union's Horizon 2020 research and innovation program (ERC Starting Grant, grant agreement No 757725).

\appendix
\section{Instanton action  and fluctuation determinants}\label{app:instanton}
The tunneling amplitude between two potential minima  can be calculated quasiclassically with the help of instanton techniques. In our model the potential is (see Fig.~\ref{fig:potential})
\begin{equation}
V(\phi)=-E_J\cos\phi-\frac{E_M}{2}\cos\frac{\phi}{2}.
\end{equation}
The main idea of this method is to find the trajectory connecting these minima that minimizes the imaginary-time action
\begin{equation}
S[\phi]=\intop_0^\beta\left(\frac{1}{16E_c}\dot{\phi}^2+V\left[\phi(\tau)\right]\right)d\tau,
\end{equation}
where $\beta=1/T$ is the inverse temperature. The action on this instanton gives the main contribution to the exponential factor of the tunneling amplitude $\nu\sim{e}^{-S_{i}}$. The trajectory $\phi_i$ is found by putting the first variation to zero,
\begin{equation}
\delta{S}=\intop_0^\beta{d}\tau\delta\phi(\tau)\left(-\frac{1}{8E_c}\dot{\phi}_{i}^2+\frac{\partial}{\partial\phi}V\left[\phi_{i}(\tau)\right]\right)=0.
\end{equation}
Then we can calculate the pre-exponent by integrating over quadratic deviations from this trajectory:
\begin{multline}
\nu=N\intop_0^\beta{d}\tau\int\,{D}\delta\phi\,\exp\left(-S_i-\frac{1}{2}\delta\phi\frac{\delta^2 S[\phi_i]}{\delta\phi^2}\delta\phi\right) \\
=\sqrt{2\pi}N\left(\mathrm{det}\,W\right)^{-1/2}e^{-S_i},
\end{multline}
where $N$ is a normalization factor, and
\begin{equation}
W=\frac{\delta^2S[\phi_i]}{\delta\phi^2}=-\frac{1}{8E_c}\frac{\partial^2}{\partial\tau^2}-\frac{\partial^2 V(\phi_i)}{\partial \phi^2}
\end{equation} is an operator that describes the fluctuations around the instanton solution, and det $W$ is the corresponding fluctuation determinant. There is always a zero mode in the spectrum of such an operator due to the fact that the instanton center $\tau_c$ can be shifted in imaginary time without changing the action. Therefore, this mode should be treated separately. Following~\cite{Vainshtein1982,Coleman1977} one can integrate over the position of an instanton center instead, which results in
\begin{equation}
\left(\mathrm{det}\,W\right)^{-1/2}=\intop_0^\beta{d}\tau_c\sqrt{\frac{S_i}{2\pi}}\left(\mathrm{det}^\prime\,W\right)^{-1/2},
\end{equation}
where $\mathrm{det}^\prime$ is the reduced determinant (with excluded zero mode). Integration over the instanton center gives the constant $\beta$.
Now we can compare the results for a non-topological junction $E_M=0$ and for a topological junction in the limit $E_M\ll{E}_J$. The operator $W$ takes the form
\begin{equation}
W=-\frac{1}{8E_c}\frac{\partial^2}{\partial\tau^2}+E_J\cos\phi+\frac{E_M}{8}\cos\frac{\phi}{2},
\end{equation}
which has a parametrically small difference between these two cases [the relative difference is $O\left(E_M/E_J\right)$]. Therefore, we can assume the reduced determinants $\mathrm{det}^\prime\,W$ for the two cases to be the same, the only significant difference arises from the zero mode, as its contribution is proportional to $\sqrt{S_i}$. This results in Eq.~(\ref{eq:nu_4pi}).
\bibliographystyle{apsrev4-1}
\bibliography{PS}

\begin{thebibliography}{60}%
\makeatletter
\providecommand \@ifxundefined [1]{%
 \@ifx{#1\undefined}
}%
\providecommand \@ifnum [1]{%
 \ifnum #1\expandafter \@firstoftwo
 \else \expandafter \@secondoftwo
 \fi
}%
\providecommand \@ifx [1]{%
 \ifx #1\expandafter \@firstoftwo
 \else \expandafter \@secondoftwo
 \fi
}%
\providecommand \natexlab [1]{#1}%
\providecommand \enquote  [1]{``#1''}%
\providecommand \bibnamefont  [1]{#1}%
\providecommand \bibfnamefont [1]{#1}%
\providecommand \citenamefont [1]{#1}%
\providecommand \href@noop [0]{\@secondoftwo}%
\providecommand \href [0]{\begingroup \@sanitize@url \@href}%
\providecommand \@href[1]{\@@startlink{#1}\@@href}%
\providecommand \@@href[1]{\endgroup#1\@@endlink}%
\providecommand \@sanitize@url [0]{\catcode `\\12\catcode `\$12\catcode
  `\&12\catcode `\#12\catcode `\^12\catcode `\_12\catcode `\%12\relax}%
\providecommand \@@startlink[1]{}%
\providecommand \@@endlink[0]{}%
\providecommand \url  [0]{\begingroup\@sanitize@url \@url }%
\providecommand \@url [1]{\endgroup\@href {#1}{\urlprefix }}%
\providecommand \urlprefix  [0]{URL }%
\providecommand \Eprint [0]{\href }%
\providecommand \doibase [0]{http://dx.doi.org/}%
\providecommand \selectlanguage [0]{\@gobble}%
\providecommand \bibinfo  [0]{\@secondoftwo}%
\providecommand \bibfield  [0]{\@secondoftwo}%
\providecommand \translation [1]{[#1]}%
\providecommand \BibitemOpen [0]{}%
\providecommand \bibitemStop [0]{}%
\providecommand \bibitemNoStop [0]{.\EOS\space}%
\providecommand \EOS [0]{\spacefactor3000\relax}%
\providecommand \BibitemShut  [1]{\csname bibitem#1\endcsname}%
\let\auto@bib@innerbib\@empty
\bibitem [{\citenamefont {Wilczek}(2009)}]{Wilczek2009}%
  \BibitemOpen
  \bibfield  {author} {\bibinfo {author} {\bibfnamefont {F.}~\bibnamefont
  {Wilczek}},\ }\href@noop {} {\bibfield  {journal} {\bibinfo  {journal} {Nat.
  Phys.}\ }\textbf {\bibinfo {volume} {5}},\ \bibinfo {pages} {614} (\bibinfo
  {year} {2009})}\BibitemShut {NoStop}%
\bibitem [{\citenamefont {Alicea}(2012)}]{Alicea2012}%
  \BibitemOpen
  \bibfield  {author} {\bibinfo {author} {\bibfnamefont {J.}~\bibnamefont
  {Alicea}},\ }\href@noop {} {\bibfield  {journal} {\bibinfo  {journal} {Rep.
  Prog. Phys.}\ }\textbf {\bibinfo {volume} {75}},\ \bibinfo {pages} {076501}
  (\bibinfo {year} {2012})}\BibitemShut {NoStop}%
\bibitem [{\citenamefont {Beenakker}(2013)}]{Beenakker2013}%
  \BibitemOpen
  \bibfield  {author} {\bibinfo {author} {\bibfnamefont {C.~W.~J.}\
  \bibnamefont {Beenakker}},\ }\href@noop {} {\bibfield  {journal} {\bibinfo
  {journal} {Annu. Rev. Condens. Matter Phys.}\ }\textbf {\bibinfo {volume}
  {4}},\ \bibinfo {pages} {113} (\bibinfo {year} {2013})}\BibitemShut {NoStop}%
\bibitem [{\citenamefont {Kitaev}(2001)}]{Kitaev2001}%
  \BibitemOpen
  \bibfield  {author} {\bibinfo {author} {\bibfnamefont {A.~Y.}\ \bibnamefont
  {Kitaev}},\ }\href@noop {} {\bibfield  {journal} {\bibinfo  {journal}
  {Physics-Uspekhi}\ }\textbf {\bibinfo {volume} {44}},\ \bibinfo {pages} {131}
  (\bibinfo {year} {2001})}\BibitemShut {NoStop}%
\bibitem [{\citenamefont {Nayak}\ \emph {et~al.}(2008)\citenamefont {Nayak},
  \citenamefont {Simon}, \citenamefont {Stern}, \citenamefont {Freedman},\ and\
  \citenamefont {Das~Sarma}}]{Nayak2008}%
  \BibitemOpen
  \bibfield  {author} {\bibinfo {author} {\bibfnamefont {C.}~\bibnamefont
  {Nayak}}, \bibinfo {author} {\bibfnamefont {S.~H.}\ \bibnamefont {Simon}},
  \bibinfo {author} {\bibfnamefont {A.}~\bibnamefont {Stern}}, \bibinfo
  {author} {\bibfnamefont {M.}~\bibnamefont {Freedman}}, \ and\ \bibinfo
  {author} {\bibfnamefont {S.}~\bibnamefont {Das~Sarma}},\ }\href@noop {}
  {\bibfield  {journal} {\bibinfo  {journal} {Rev. Mod. Phys.}\ }\textbf
  {\bibinfo {volume} {80}},\ \bibinfo {pages} {1083} (\bibinfo {year}
  {2008})}\BibitemShut {NoStop}%
\bibitem [{\citenamefont {Alicea}\ \emph {et~al.}(2011)\citenamefont {Alicea},
  \citenamefont {Oreg}, \citenamefont {Refael}, \citenamefont {von Oppen},\
  and\ \citenamefont {Fisher}}]{Alicea2011}%
  \BibitemOpen
  \bibfield  {author} {\bibinfo {author} {\bibfnamefont {J.}~\bibnamefont
  {Alicea}}, \bibinfo {author} {\bibfnamefont {Y.}~\bibnamefont {Oreg}},
  \bibinfo {author} {\bibfnamefont {G.}~\bibnamefont {Refael}}, \bibinfo
  {author} {\bibfnamefont {F.}~\bibnamefont {von Oppen}}, \ and\ \bibinfo
  {author} {\bibfnamefont {M.~P.~A.}\ \bibnamefont {Fisher}},\ }\href@noop {}
  {\bibfield  {journal} {\bibinfo  {journal} {Nat. Phys.}\ }\textbf {\bibinfo
  {volume} {7}},\ \bibinfo {pages} {412} (\bibinfo {year} {2011})}\BibitemShut
  {NoStop}%
\bibitem [{\citenamefont {Hoffman}\ \emph {et~al.}(2016)\citenamefont
  {Hoffman}, \citenamefont {Schrade}, \citenamefont {Klinovaja},\ and\
  \citenamefont {Loss}}]{Hoffman2016}%
  \BibitemOpen
  \bibfield  {author} {\bibinfo {author} {\bibfnamefont {S.}~\bibnamefont
  {Hoffman}}, \bibinfo {author} {\bibfnamefont {C.}~\bibnamefont {Schrade}},
  \bibinfo {author} {\bibfnamefont {J.}~\bibnamefont {Klinovaja}}, \ and\
  \bibinfo {author} {\bibfnamefont {D.}~\bibnamefont {Loss}},\ }\href@noop {}
  {\bibfield  {journal} {\bibinfo  {journal} {Phys. Rev. B}\ }\textbf {\bibinfo
  {volume} {94}},\ \bibinfo {pages} {045316} (\bibinfo {year}
  {2016})}\BibitemShut {NoStop}%
\bibitem [{\citenamefont {Plugge}\ \emph {et~al.}(2017)\citenamefont {Plugge},
  \citenamefont {Rasmussen}, \citenamefont {Egger},\ and\ \citenamefont
  {Flensberg}}]{Plugge2017}%
  \BibitemOpen
  \bibfield  {author} {\bibinfo {author} {\bibfnamefont {S.}~\bibnamefont
  {Plugge}}, \bibinfo {author} {\bibfnamefont {A.}~\bibnamefont {Rasmussen}},
  \bibinfo {author} {\bibfnamefont {R.}~\bibnamefont {Egger}}, \ and\ \bibinfo
  {author} {\bibfnamefont {K.}~\bibnamefont {Flensberg}},\ }\href@noop {}
  {\bibfield  {journal} {\bibinfo  {journal} {New. J. Phys.}\ }\textbf
  {\bibinfo {volume} {19}},\ \bibinfo {pages} {012001} (\bibinfo {year}
  {2017})}\BibitemShut {NoStop}%
\bibitem [{\citenamefont {Woods}\ \emph {et~al.}(2019)\citenamefont {Woods},
  \citenamefont {Chen}, \citenamefont {Frolov},\ and\ \citenamefont
  {Stanescu}}]{Stanescu2019}%
  \BibitemOpen
  \bibfield  {author} {\bibinfo {author} {\bibfnamefont {B.~D.}\ \bibnamefont
  {Woods}}, \bibinfo {author} {\bibfnamefont {J.}~\bibnamefont {Chen}},
  \bibinfo {author} {\bibfnamefont {S.~M.}\ \bibnamefont {Frolov}}, \ and\
  \bibinfo {author} {\bibfnamefont {T.~D.}\ \bibnamefont {Stanescu}},\
  }\href@noop {} {\bibfield  {journal} {\bibinfo  {journal} {Phys. Rev. B}\
  }\textbf {\bibinfo {volume} {100}},\ \bibinfo {pages} {125407} (\bibinfo
  {year} {2019})}\BibitemShut {NoStop}%
\bibitem [{\citenamefont {Reeg}\ \emph {et~al.}(2018)\citenamefont {Reeg},
  \citenamefont {Dmytruk}, \citenamefont {Chevallier}, \citenamefont {Loss},\
  and\ \citenamefont {Klinovaja}}]{Reeg2018}%
  \BibitemOpen
  \bibfield  {author} {\bibinfo {author} {\bibfnamefont {C.}~\bibnamefont
  {Reeg}}, \bibinfo {author} {\bibfnamefont {O.}~\bibnamefont {Dmytruk}},
  \bibinfo {author} {\bibfnamefont {D.}~\bibnamefont {Chevallier}}, \bibinfo
  {author} {\bibfnamefont {D.}~\bibnamefont {Loss}}, \ and\ \bibinfo {author}
  {\bibfnamefont {J.}~\bibnamefont {Klinovaja}},\ }\href@noop {} {\bibfield
  {journal} {\bibinfo  {journal} {Phys. Rev. B}\ }\textbf {\bibinfo {volume}
  {98}},\ \bibinfo {pages} {245407} (\bibinfo {year} {2018})}\BibitemShut
  {NoStop}%
\bibitem [{\citenamefont {Yu}\ \emph {et~al.}(2020)\citenamefont {Yu},
  \citenamefont {Chen}, \citenamefont {Gomanko}, \citenamefont {Badawy},
  \citenamefont {Bakkers}, \citenamefont {Zuo}, \citenamefont {Mourik},\ and\
  \citenamefont {Frolov}}]{Frolov2020}%
  \BibitemOpen
  \bibfield  {author} {\bibinfo {author} {\bibfnamefont {P.}~\bibnamefont
  {Yu}}, \bibinfo {author} {\bibfnamefont {J.}~\bibnamefont {Chen}}, \bibinfo
  {author} {\bibfnamefont {M.}~\bibnamefont {Gomanko}}, \bibinfo {author}
  {\bibfnamefont {G.}~\bibnamefont {Badawy}}, \bibinfo {author} {\bibfnamefont
  {E.~P. A.~M.}\ \bibnamefont {Bakkers}}, \bibinfo {author} {\bibfnamefont
  {K.}~\bibnamefont {Zuo}}, \bibinfo {author} {\bibfnamefont {V.}~\bibnamefont
  {Mourik}}, \ and\ \bibinfo {author} {\bibfnamefont {S.~M.}\ \bibnamefont
  {Frolov}},\ }\href@noop {} {\bibfield  {journal} {\bibinfo  {journal}
  {arXiv:2004.08583}\ } (\bibinfo {year} {2020})}\BibitemShut {NoStop}%
\bibitem [{\citenamefont {Liu}\ \emph {et~al.}(2018)\citenamefont {Liu},
  \citenamefont {Sau},\ and\ \citenamefont {Das~Sarma}}]{Sarma2018}%
  \BibitemOpen
  \bibfield  {author} {\bibinfo {author} {\bibfnamefont {C.-X.}\ \bibnamefont
  {Liu}}, \bibinfo {author} {\bibfnamefont {J.~D.}\ \bibnamefont {Sau}}, \ and\
  \bibinfo {author} {\bibfnamefont {S.}~\bibnamefont {Das~Sarma}},\ }\href@noop
  {} {\bibfield  {journal} {\bibinfo  {journal} {Phys. Rev. B}\ }\textbf
  {\bibinfo {volume} {97}},\ \bibinfo {pages} {214502} (\bibinfo {year}
  {2018})}\BibitemShut {NoStop}%
\bibitem [{\citenamefont {Moore}\ \emph {et~al.}(2018)\citenamefont {Moore},
  \citenamefont {Zeng}, \citenamefont {Stanescu},\ and\ \citenamefont
  {Tewari}}]{Moore2018}%
  \BibitemOpen
  \bibfield  {author} {\bibinfo {author} {\bibfnamefont {C.}~\bibnamefont
  {Moore}}, \bibinfo {author} {\bibfnamefont {C.}~\bibnamefont {Zeng}},
  \bibinfo {author} {\bibfnamefont {T.~D.}\ \bibnamefont {Stanescu}}, \ and\
  \bibinfo {author} {\bibfnamefont {S.}~\bibnamefont {Tewari}},\ }\href@noop {}
  {\bibfield  {journal} {\bibinfo  {journal} {Phys. Rev. B}\ }\textbf {\bibinfo
  {volume} {98}},\ \bibinfo {pages} {155314} (\bibinfo {year}
  {2018})}\BibitemShut {NoStop}%
\bibitem [{\citenamefont {Peñaranda}\ \emph {et~al.}(2018)\citenamefont
  {Peñaranda}, \citenamefont {Aguado}, \citenamefont {San-Jose},\ and\
  \citenamefont {Prada}}]{Prada2018}%
  \BibitemOpen
  \bibfield  {author} {\bibinfo {author} {\bibfnamefont {F.}~\bibnamefont
  {Peñaranda}}, \bibinfo {author} {\bibfnamefont {R.}~\bibnamefont {Aguado}},
  \bibinfo {author} {\bibfnamefont {P.}~\bibnamefont {San-Jose}}, \ and\
  \bibinfo {author} {\bibfnamefont {E.}~\bibnamefont {Prada}},\ }\href@noop {}
  {\bibfield  {journal} {\bibinfo  {journal} {Phys. Rev. B}\ }\textbf {\bibinfo
  {volume} {98}},\ \bibinfo {pages} {235406} (\bibinfo {year}
  {2018})}\BibitemShut {NoStop}%
\bibitem [{\citenamefont {Prada}\ \emph {et~al.}(2019)\citenamefont {Prada},
  \citenamefont {San-Jose}, \citenamefont {de~Moor}, \citenamefont {Geresdi},
  \citenamefont {Lee}, \citenamefont {Klinovaja}, \citenamefont {Loss},
  \citenamefont {Nygård}, \citenamefont {Aguado},\ and\ \citenamefont
  {Kouwenhoven}}]{Prada2019}%
  \BibitemOpen
  \bibfield  {author} {\bibinfo {author} {\bibfnamefont {E.}~\bibnamefont
  {Prada}}, \bibinfo {author} {\bibfnamefont {P.}~\bibnamefont {San-Jose}},
  \bibinfo {author} {\bibfnamefont {M.~W.~A.}\ \bibnamefont {de~Moor}},
  \bibinfo {author} {\bibfnamefont {A.}~\bibnamefont {Geresdi}}, \bibinfo
  {author} {\bibfnamefont {E.~J.~H.}\ \bibnamefont {Lee}}, \bibinfo {author}
  {\bibfnamefont {J.}~\bibnamefont {Klinovaja}}, \bibinfo {author}
  {\bibfnamefont {D.}~\bibnamefont {Loss}}, \bibinfo {author} {\bibfnamefont
  {J.}~\bibnamefont {Nygård}}, \bibinfo {author} {\bibfnamefont
  {R.}~\bibnamefont {Aguado}}, \ and\ \bibinfo {author} {\bibfnamefont {L.~P.}\
  \bibnamefont {Kouwenhoven}},\ }\href@noop {} {\bibfield  {journal} {\bibinfo
  {journal} {arXiv:1911.04512}\ } (\bibinfo {year} {2019})}\BibitemShut
  {NoStop}%
\bibitem [{\citenamefont {Ioselevich}\ and\ \citenamefont
  {Feigel’man}(2011)}]{Ioselevich2011}%
  \BibitemOpen
  \bibfield  {author} {\bibinfo {author} {\bibfnamefont {P.~A.}\ \bibnamefont
  {Ioselevich}}\ and\ \bibinfo {author} {\bibfnamefont {M.~V.}\ \bibnamefont
  {Feigel’man}},\ }\href@noop {} {\bibfield  {journal} {\bibinfo  {journal}
  {Phys. Rev. Lett}\ }\textbf {\bibinfo {volume} {106}},\ \bibinfo {pages}
  {077003} (\bibinfo {year} {2011})}\BibitemShut {NoStop}%
\bibitem [{\citenamefont {Kurter}\ \emph {et~al.}(2015)\citenamefont {Kurter},
  \citenamefont {Finck}, \citenamefont {Hor},\ and\ \citenamefont
  {Van~Harlingen}}]{Kurter2015}%
  \BibitemOpen
  \bibfield  {author} {\bibinfo {author} {\bibfnamefont {C.}~\bibnamefont
  {Kurter}}, \bibinfo {author} {\bibfnamefont {A.}~\bibnamefont {Finck}},
  \bibinfo {author} {\bibfnamefont {Y.~S.}\ \bibnamefont {Hor}}, \ and\
  \bibinfo {author} {\bibfnamefont {D.~J.}\ \bibnamefont {Van~Harlingen}},\
  }\href@noop {} {\bibfield  {journal} {\bibinfo  {journal} {Nat. Comm.}\
  }\textbf {\bibinfo {volume} {6}},\ \bibinfo {pages} {7130} (\bibinfo {year}
  {2015})}\BibitemShut {NoStop}%
\bibitem [{\citenamefont {Wiedenmann}\ \emph {et~al.}(2016)\citenamefont
  {Wiedenmann}, \citenamefont {Bocquillon}, \citenamefont {Deacon},
  \citenamefont {Hartinger}, \citenamefont {Herrmann}, \citenamefont
  {Klapwijk}, \citenamefont {Maier}, \citenamefont {Ames}, \citenamefont
  {Brüne}, \citenamefont {Gould}, \citenamefont {Oiwa}, \citenamefont
  {Ishibashi}, \citenamefont {Tarucha}, \citenamefont {Buhmann},\ and\
  \citenamefont {Molenkamp}}]{Wiedenmann2016}%
  \BibitemOpen
  \bibfield  {author} {\bibinfo {author} {\bibfnamefont {J.}~\bibnamefont
  {Wiedenmann}}, \bibinfo {author} {\bibfnamefont {E.}~\bibnamefont
  {Bocquillon}}, \bibinfo {author} {\bibfnamefont {R.~S.}\ \bibnamefont
  {Deacon}}, \bibinfo {author} {\bibfnamefont {S.}~\bibnamefont {Hartinger}},
  \bibinfo {author} {\bibfnamefont {O.}~\bibnamefont {Herrmann}}, \bibinfo
  {author} {\bibfnamefont {T.~M.}\ \bibnamefont {Klapwijk}}, \bibinfo {author}
  {\bibfnamefont {L.}~\bibnamefont {Maier}}, \bibinfo {author} {\bibfnamefont
  {C.}~\bibnamefont {Ames}}, \bibinfo {author} {\bibfnamefont {C.}~\bibnamefont
  {Brüne}}, \bibinfo {author} {\bibfnamefont {C.}~\bibnamefont {Gould}},
  \bibinfo {author} {\bibfnamefont {A.}~\bibnamefont {Oiwa}}, \bibinfo {author}
  {\bibfnamefont {K.}~\bibnamefont {Ishibashi}}, \bibinfo {author}
  {\bibfnamefont {S.}~\bibnamefont {Tarucha}}, \bibinfo {author} {\bibfnamefont
  {H.}~\bibnamefont {Buhmann}}, \ and\ \bibinfo {author} {\bibfnamefont
  {L.~W.}\ \bibnamefont {Molenkamp}},\ }\href@noop {} {\bibfield  {journal}
  {\bibinfo  {journal} {Nat. Comm.}\ }\textbf {\bibinfo {volume} {7}},\
  \bibinfo {pages} {10303} (\bibinfo {year} {2016})}\BibitemShut {NoStop}%
\bibitem [{\citenamefont {Oreg}\ \emph {et~al.}(2010)\citenamefont {Oreg},
  \citenamefont {Refael},\ and\ \citenamefont {von Oppen}}]{Oreg2010}%
  \BibitemOpen
  \bibfield  {author} {\bibinfo {author} {\bibfnamefont {Y.}~\bibnamefont
  {Oreg}}, \bibinfo {author} {\bibfnamefont {G.}~\bibnamefont {Refael}}, \ and\
  \bibinfo {author} {\bibfnamefont {F.}~\bibnamefont {von Oppen}},\ }\href@noop
  {} {\bibfield  {journal} {\bibinfo  {journal} {Phys. Rev. Lett.}\ }\textbf
  {\bibinfo {volume} {105}},\ \bibinfo {pages} {177002} (\bibinfo {year}
  {2010})}\BibitemShut {NoStop}%
\bibitem [{\citenamefont {Lutchyn}\ \emph {et~al.}(2010)\citenamefont
  {Lutchyn}, \citenamefont {Sau},\ and\ \citenamefont {Das~Sarma}}]{Sau2010}%
  \BibitemOpen
  \bibfield  {author} {\bibinfo {author} {\bibfnamefont {R.~M.}\ \bibnamefont
  {Lutchyn}}, \bibinfo {author} {\bibfnamefont {J.~D.}\ \bibnamefont {Sau}}, \
  and\ \bibinfo {author} {\bibfnamefont {S.}~\bibnamefont {Das~Sarma}},\
  }\href@noop {} {\bibfield  {journal} {\bibinfo  {journal} {Phys. Rev. Lett.}\
  }\textbf {\bibinfo {volume} {105}},\ \bibinfo {pages} {077001} (\bibinfo
  {year} {2010})}\BibitemShut {NoStop}%
\bibitem [{\citenamefont {San-Jose}\ \emph {et~al.}(2013)\citenamefont
  {San-Jose}, \citenamefont {Cayao}, \citenamefont {Prada},\ and\ \citenamefont
  {Aguado}}]{San-Jose2013}%
  \BibitemOpen
  \bibfield  {author} {\bibinfo {author} {\bibfnamefont {P.}~\bibnamefont
  {San-Jose}}, \bibinfo {author} {\bibfnamefont {J.}~\bibnamefont {Cayao}},
  \bibinfo {author} {\bibfnamefont {E.}~\bibnamefont {Prada}}, \ and\ \bibinfo
  {author} {\bibfnamefont {R.}~\bibnamefont {Aguado}},\ }\href@noop {}
  {\bibfield  {journal} {\bibinfo  {journal} {New J. Phys.}\ }\textbf {\bibinfo
  {volume} {15}},\ \bibinfo {pages} {075019} (\bibinfo {year}
  {2013})}\BibitemShut {NoStop}%
\bibitem [{\citenamefont {Albrecht}\ \emph {et~al.}(2016)\citenamefont
  {Albrecht}, \citenamefont {Higginbotham}, \citenamefont {Madsen},
  \citenamefont {Kuemmeth}, \citenamefont {Jespersen}, \citenamefont
  {Krogstrup},\ and\ \citenamefont {Marcus}}]{Albrecht2016}%
  \BibitemOpen
  \bibfield  {author} {\bibinfo {author} {\bibfnamefont {S.~M.}\ \bibnamefont
  {Albrecht}}, \bibinfo {author} {\bibfnamefont {A.~P.}\ \bibnamefont
  {Higginbotham}}, \bibinfo {author} {\bibfnamefont {M.}~\bibnamefont
  {Madsen}}, \bibinfo {author} {\bibfnamefont {F.}~\bibnamefont {Kuemmeth}},
  \bibinfo {author} {\bibfnamefont {J.}~\bibnamefont {Jespersen}, \bibfnamefont
  {T.~S.~Nygård}}, \bibinfo {author} {\bibfnamefont {P.}~\bibnamefont
  {Krogstrup}}, \ and\ \bibinfo {author} {\bibfnamefont {C.~M.}\ \bibnamefont
  {Marcus}},\ }\href@noop {} {\bibfield  {journal} {\bibinfo  {journal}
  {Nature}\ }\textbf {\bibinfo {volume} {531}},\ \bibinfo {pages} {206}
  (\bibinfo {year} {2016})}\BibitemShut {NoStop}%
\bibitem [{\citenamefont {Nesterov}\ \emph {et~al.}(2016)\citenamefont
  {Nesterov}, \citenamefont {Houzet},\ and\ \citenamefont
  {Meyer}}]{Nesterov2016}%
  \BibitemOpen
  \bibfield  {author} {\bibinfo {author} {\bibfnamefont {K.~N.}\ \bibnamefont
  {Nesterov}}, \bibinfo {author} {\bibfnamefont {M.}~\bibnamefont {Houzet}}, \
  and\ \bibinfo {author} {\bibfnamefont {J.~S.}\ \bibnamefont {Meyer}},\
  }\href@noop {} {\bibfield  {journal} {\bibinfo  {journal} {Phys. Rev. B}\
  }\textbf {\bibinfo {volume} {93}},\ \bibinfo {pages} {174502} (\bibinfo
  {year} {2016})}\BibitemShut {NoStop}%
\bibitem [{\citenamefont {de~Moor}\ \emph {et~al.}(2018)\citenamefont
  {de~Moor}, \citenamefont {Bommer}, \citenamefont {Xu}, \citenamefont
  {Winkler}, \citenamefont {Antipov}, \citenamefont {Bargerbos}, \citenamefont
  {Wang}, \citenamefont {van Loo}, \citenamefont {Op~het Veld}, \citenamefont
  {Gazibegovic}, \citenamefont {Car}, \citenamefont {Logan}, \citenamefont
  {Pendharkar}, \citenamefont {Lee}, \citenamefont {Bakkers}, \citenamefont
  {Palmstrøm}, \citenamefont {Lutchyn}, \citenamefont {Kouwenhoven},\ and\
  \citenamefont {Zhang}}]{deMoor2018}%
  \BibitemOpen
  \bibfield  {author} {\bibinfo {author} {\bibfnamefont {M.~W.~A.}\
  \bibnamefont {de~Moor}}, \bibinfo {author} {\bibfnamefont {J.~D.~S.}\
  \bibnamefont {Bommer}}, \bibinfo {author} {\bibfnamefont {D.}~\bibnamefont
  {Xu}}, \bibinfo {author} {\bibfnamefont {G.~W.}\ \bibnamefont {Winkler}},
  \bibinfo {author} {\bibfnamefont {A.~E.}\ \bibnamefont {Antipov}}, \bibinfo
  {author} {\bibfnamefont {A.}~\bibnamefont {Bargerbos}}, \bibinfo {author}
  {\bibfnamefont {G.}~\bibnamefont {Wang}}, \bibinfo {author} {\bibfnamefont
  {N.}~\bibnamefont {van Loo}}, \bibinfo {author} {\bibfnamefont {R.~L.~M.}\
  \bibnamefont {Op~het Veld}}, \bibinfo {author} {\bibfnamefont
  {S.}~\bibnamefont {Gazibegovic}}, \bibinfo {author} {\bibfnamefont
  {D.}~\bibnamefont {Car}}, \bibinfo {author} {\bibfnamefont {J.~A.}\
  \bibnamefont {Logan}}, \bibinfo {author} {\bibfnamefont {M.}~\bibnamefont
  {Pendharkar}}, \bibinfo {author} {\bibfnamefont {J.~S.}\ \bibnamefont {Lee}},
  \bibinfo {author} {\bibfnamefont {E.~P. A.~M.}\ \bibnamefont {Bakkers}},
  \bibinfo {author} {\bibfnamefont {C.~J.}\ \bibnamefont {Palmstrøm}},
  \bibinfo {author} {\bibfnamefont {R.~M.}\ \bibnamefont {Lutchyn}}, \bibinfo
  {author} {\bibfnamefont {L.~P.}\ \bibnamefont {Kouwenhoven}}, \ and\ \bibinfo
  {author} {\bibfnamefont {H.}~\bibnamefont {Zhang}},\ }\href@noop {}
  {\bibfield  {journal} {\bibinfo  {journal} {New J. Phys.}\ }\textbf {\bibinfo
  {volume} {20}},\ \bibinfo {pages} {103049} (\bibinfo {year}
  {2018})}\BibitemShut {NoStop}%
\bibitem [{\citenamefont {Schrade}\ \emph {et~al.}(2017)\citenamefont
  {Schrade}, \citenamefont {Hoffman},\ and\ \citenamefont
  {Loss}}]{Schrade2017}%
  \BibitemOpen
  \bibfield  {author} {\bibinfo {author} {\bibfnamefont {C.}~\bibnamefont
  {Schrade}}, \bibinfo {author} {\bibfnamefont {S.}~\bibnamefont {Hoffman}}, \
  and\ \bibinfo {author} {\bibfnamefont {D.}~\bibnamefont {Loss}},\ }\href@noop
  {} {\bibfield  {journal} {\bibinfo  {journal} {Phys. Rev. B}\ }\textbf
  {\bibinfo {volume} {95}},\ \bibinfo {pages} {195421} (\bibinfo {year}
  {2017})}\BibitemShut {NoStop}%
\bibitem [{\citenamefont {Dolcini}\ \emph {et~al.}(2015)\citenamefont
  {Dolcini}, \citenamefont {Houzet},\ and\ \citenamefont
  {Meyer}}]{Dolcini2015}%
  \BibitemOpen
  \bibfield  {author} {\bibinfo {author} {\bibfnamefont {F.}~\bibnamefont
  {Dolcini}}, \bibinfo {author} {\bibfnamefont {M.}~\bibnamefont {Houzet}}, \
  and\ \bibinfo {author} {\bibfnamefont {J.~S.}\ \bibnamefont {Meyer}},\
  }\href@noop {} {\bibfield  {journal} {\bibinfo  {journal} {Phys. Rev. B}\
  }\textbf {\bibinfo {volume} {92}},\ \bibinfo {pages} {035428} (\bibinfo
  {year} {2015})}\BibitemShut {NoStop}%
\bibitem [{\citenamefont {Klinovaja}\ \emph {et~al.}(2012)\citenamefont
  {Klinovaja}, \citenamefont {Gangadharaiah},\ and\ \citenamefont
  {Loss}}]{Klinovaja2012}%
  \BibitemOpen
  \bibfield  {author} {\bibinfo {author} {\bibfnamefont {J.}~\bibnamefont
  {Klinovaja}}, \bibinfo {author} {\bibfnamefont {S.}~\bibnamefont
  {Gangadharaiah}}, \ and\ \bibinfo {author} {\bibfnamefont {D.}~\bibnamefont
  {Loss}},\ }\href@noop {} {\bibfield  {journal} {\bibinfo  {journal} {Phys.
  Rev. Lett.}\ }\textbf {\bibinfo {volume} {108}},\ \bibinfo {pages} {196804}
  (\bibinfo {year} {2012})}\BibitemShut {NoStop}%
\bibitem [{\citenamefont {Sau}\ and\ \citenamefont {Tewari}(2013)}]{Sau2013}%
  \BibitemOpen
  \bibfield  {author} {\bibinfo {author} {\bibfnamefont {J.~D.}\ \bibnamefont
  {Sau}}\ and\ \bibinfo {author} {\bibfnamefont {S.}~\bibnamefont {Tewari}},\
  }\href@noop {} {\bibfield  {journal} {\bibinfo  {journal} {Phys. Rev. B}\
  }\textbf {\bibinfo {volume} {88}},\ \bibinfo {pages} {054503} (\bibinfo
  {year} {2013})}\BibitemShut {NoStop}%
\bibitem [{\citenamefont {Marganska}\ \emph {et~al.}(2018)\citenamefont
  {Marganska}, \citenamefont {Milz}, \citenamefont {Izumida}, \citenamefont
  {Strunk},\ and\ \citenamefont {Grifoni}}]{Marganska2018}%
  \BibitemOpen
  \bibfield  {author} {\bibinfo {author} {\bibfnamefont {M.}~\bibnamefont
  {Marganska}}, \bibinfo {author} {\bibfnamefont {L.}~\bibnamefont {Milz}},
  \bibinfo {author} {\bibfnamefont {W.}~\bibnamefont {Izumida}}, \bibinfo
  {author} {\bibfnamefont {C.}~\bibnamefont {Strunk}}, \ and\ \bibinfo {author}
  {\bibfnamefont {M.}~\bibnamefont {Grifoni}},\ }\href@noop {} {\bibfield
  {journal} {\bibinfo  {journal} {Phys. Rev. B}\ }\textbf {\bibinfo {volume}
  {97}},\ \bibinfo {pages} {075141} (\bibinfo {year} {2018})}\BibitemShut
  {NoStop}%
\bibitem [{\citenamefont {Alicea}(2010)}]{Alicea2010}%
  \BibitemOpen
  \bibfield  {author} {\bibinfo {author} {\bibfnamefont {J.}~\bibnamefont
  {Alicea}},\ }\href@noop {} {\bibfield  {journal} {\bibinfo  {journal} {Phys.
  Rev. B}\ }\textbf {\bibinfo {volume} {81}},\ \bibinfo {pages} {125318}
  (\bibinfo {year} {2010})}\BibitemShut {NoStop}%
\bibitem [{\citenamefont {Krogstrup}\ \emph {et~al.}(2015)\citenamefont
  {Krogstrup}, \citenamefont {Ziino}, \citenamefont {Chang}, \citenamefont
  {Albrecht}, \citenamefont {Madsen}, \citenamefont {Johnson}, \citenamefont
  {Nygård}, \citenamefont {Marcus},\ and\ \citenamefont
  {Jespersen}}]{Krogstrup2015}%
  \BibitemOpen
  \bibfield  {author} {\bibinfo {author} {\bibfnamefont {P.}~\bibnamefont
  {Krogstrup}}, \bibinfo {author} {\bibfnamefont {N.~L.~B.}\ \bibnamefont
  {Ziino}}, \bibinfo {author} {\bibfnamefont {W.}~\bibnamefont {Chang}},
  \bibinfo {author} {\bibfnamefont {S.~M.}\ \bibnamefont {Albrecht}}, \bibinfo
  {author} {\bibfnamefont {M.~H.}\ \bibnamefont {Madsen}}, \bibinfo {author}
  {\bibfnamefont {E.}~\bibnamefont {Johnson}}, \bibinfo {author} {\bibfnamefont
  {J.}~\bibnamefont {Nygård}}, \bibinfo {author} {\bibfnamefont {C.~M.}\
  \bibnamefont {Marcus}}, \ and\ \bibinfo {author} {\bibfnamefont {T.~S.}\
  \bibnamefont {Jespersen}},\ }\href@noop {} {\bibfield  {journal} {\bibinfo
  {journal} {Nat. Mater.}\ }\textbf {\bibinfo {volume} {14}},\ \bibinfo {pages}
  {400} (\bibinfo {year} {2015})}\BibitemShut {NoStop}%
\bibitem [{\citenamefont {Deng}\ \emph {et~al.}(2016)\citenamefont {Deng},
  \citenamefont {Vaitiekenas}, \citenamefont {Hansen}, \citenamefont {Danon},
  \citenamefont {Leijnse}, \citenamefont {Flensberg}, \citenamefont {Nygård},
  \citenamefont {Krogstrup},\ and\ \citenamefont {Marcus}}]{Deng2016}%
  \BibitemOpen
  \bibfield  {author} {\bibinfo {author} {\bibfnamefont {M.~T.}\ \bibnamefont
  {Deng}}, \bibinfo {author} {\bibfnamefont {S.}~\bibnamefont {Vaitiekenas}},
  \bibinfo {author} {\bibfnamefont {E.~B.}\ \bibnamefont {Hansen}}, \bibinfo
  {author} {\bibfnamefont {J.}~\bibnamefont {Danon}}, \bibinfo {author}
  {\bibfnamefont {M.}~\bibnamefont {Leijnse}}, \bibinfo {author} {\bibfnamefont
  {K.}~\bibnamefont {Flensberg}}, \bibinfo {author} {\bibfnamefont
  {J.}~\bibnamefont {Nygård}}, \bibinfo {author} {\bibfnamefont
  {P.}~\bibnamefont {Krogstrup}}, \ and\ \bibinfo {author} {\bibfnamefont
  {C.~M.}\ \bibnamefont {Marcus}},\ }\href@noop {} {\bibfield  {journal}
  {\bibinfo  {journal} {Science}\ }\textbf {\bibinfo {volume} {354}},\ \bibinfo
  {pages} {1557} (\bibinfo {year} {2016})}\BibitemShut {NoStop}%
\bibitem [{\citenamefont {Kwon}\ \emph {et~al.}(2003)\citenamefont {Kwon},
  \citenamefont {Sengupta},\ and\ \citenamefont {Yakovenko}}]{Kwon2003}%
  \BibitemOpen
  \bibfield  {author} {\bibinfo {author} {\bibfnamefont {H.-J.}\ \bibnamefont
  {Kwon}}, \bibinfo {author} {\bibfnamefont {K.}~\bibnamefont {Sengupta}}, \
  and\ \bibinfo {author} {\bibfnamefont {V.~M.}\ \bibnamefont {Yakovenko}},\
  }\href@noop {} {\bibfield  {journal} {\bibinfo  {journal} {Eur. Phys. J. B}\
  }\textbf {\bibinfo {volume} {37}},\ \bibinfo {pages} {349} (\bibinfo {year}
  {2003})}\BibitemShut {NoStop}%
\bibitem [{\citenamefont {Fu}\ and\ \citenamefont {Kane}(2009)}]{Fu2009}%
  \BibitemOpen
  \bibfield  {author} {\bibinfo {author} {\bibfnamefont {L.}~\bibnamefont
  {Fu}}\ and\ \bibinfo {author} {\bibfnamefont {C.~L.}\ \bibnamefont {Kane}},\
  }\href@noop {} {\bibfield  {journal} {\bibinfo  {journal} {Phys. Rev. B}\
  }\textbf {\bibinfo {volume} {79}},\ \bibinfo {pages} {161408(R)} (\bibinfo
  {year} {2009})}\BibitemShut {NoStop}%
\bibitem [{\citenamefont {Pekker}\ \emph {et~al.}(2013)\citenamefont {Pekker},
  \citenamefont {Hou}, \citenamefont {Bergman}, \citenamefont {Goldberg},
  \citenamefont {Adagideli},\ and\ \citenamefont {Hassler}}]{Pekker2018}%
  \BibitemOpen
  \bibfield  {author} {\bibinfo {author} {\bibfnamefont {D.}~\bibnamefont
  {Pekker}}, \bibinfo {author} {\bibfnamefont {C.-Y.}\ \bibnamefont {Hou}},
  \bibinfo {author} {\bibfnamefont {D.~L.}\ \bibnamefont {Bergman}}, \bibinfo
  {author} {\bibfnamefont {S.}~\bibnamefont {Goldberg}}, \bibinfo {author}
  {\bibfnamefont {{\.I}.}~\bibnamefont {Adagideli}}, \ and\ \bibinfo {author}
  {\bibfnamefont {F.}~\bibnamefont {Hassler}},\ }\href@noop {} {\bibfield
  {journal} {\bibinfo  {journal} {Phys. Rev. B}\ }\textbf {\bibinfo {volume}
  {87}},\ \bibinfo {pages} {064506} (\bibinfo {year} {2013})}\BibitemShut
  {NoStop}%
\bibitem [{\citenamefont {{\'A}vila}\ \emph {et~al.}(2020)\citenamefont
  {{\'A}vila}, \citenamefont {Prada}, \citenamefont {San-Jose},\ and\
  \citenamefont {Aguado}}]{Aguado2020_1}%
  \BibitemOpen
  \bibfield  {author} {\bibinfo {author} {\bibfnamefont {J.}~\bibnamefont
  {{\'A}vila}}, \bibinfo {author} {\bibfnamefont {E.}~\bibnamefont {Prada}},
  \bibinfo {author} {\bibfnamefont {P.}~\bibnamefont {San-Jose}}, \ and\
  \bibinfo {author} {\bibfnamefont {R.}~\bibnamefont {Aguado}},\ }\href@noop {}
  {\bibfield  {journal} {\bibinfo  {journal} {arXiv:2003.02852v1}\ } (\bibinfo
  {year} {2020})}\BibitemShut {NoStop}%
\bibitem [{\citenamefont {Likharev}\ and\ \citenamefont
  {Zorin}(1985)}]{Likharev1985}%
  \BibitemOpen
  \bibfield  {author} {\bibinfo {author} {\bibfnamefont {K.~K.}\ \bibnamefont
  {Likharev}}\ and\ \bibinfo {author} {\bibfnamefont {A.~B.}\ \bibnamefont
  {Zorin}},\ }\href@noop {} {\bibfield  {journal} {\bibinfo  {journal} {J. of
  Low Temp. Phys.}\ }\textbf {\bibinfo {volume} {59}},\ \bibinfo {pages} {347}
  (\bibinfo {year} {1985})}\BibitemShut {NoStop}%
\bibitem [{\citenamefont {Jiang}\ \emph {et~al.}(2011)\citenamefont {Jiang},
  \citenamefont {Pekker}, \citenamefont {Alicea}, \citenamefont {Refael},
  \citenamefont {Oreg},\ and\ \citenamefont {von Oppen}}]{Oppen2011}%
  \BibitemOpen
  \bibfield  {author} {\bibinfo {author} {\bibfnamefont {L.}~\bibnamefont
  {Jiang}}, \bibinfo {author} {\bibfnamefont {D.}~\bibnamefont {Pekker}},
  \bibinfo {author} {\bibfnamefont {J.}~\bibnamefont {Alicea}}, \bibinfo
  {author} {\bibfnamefont {G.}~\bibnamefont {Refael}}, \bibinfo {author}
  {\bibfnamefont {Y.}~\bibnamefont {Oreg}}, \ and\ \bibinfo {author}
  {\bibfnamefont {F.}~\bibnamefont {von Oppen}},\ }\href@noop {} {\bibfield
  {journal} {\bibinfo  {journal} {Phys. Rev. Lett}\ }\textbf {\bibinfo {volume}
  {107}},\ \bibinfo {pages} {236401} (\bibinfo {year} {2011})}\BibitemShut
  {NoStop}%
\bibitem [{\citenamefont {Zyuzin}\ \emph {et~al.}(2013)\citenamefont {Zyuzin},
  \citenamefont {Rainis}, \citenamefont {Klinovaja},\ and\ \citenamefont
  {Loss}}]{Zyuzin2013}%
  \BibitemOpen
  \bibfield  {author} {\bibinfo {author} {\bibfnamefont {A.~A.}\ \bibnamefont
  {Zyuzin}}, \bibinfo {author} {\bibfnamefont {D.}~\bibnamefont {Rainis}},
  \bibinfo {author} {\bibfnamefont {J.}~\bibnamefont {Klinovaja}}, \ and\
  \bibinfo {author} {\bibfnamefont {D.}~\bibnamefont {Loss}},\ }\href@noop {}
  {\bibfield  {journal} {\bibinfo  {journal} {Phys. Rev. Lett.}\ }\textbf
  {\bibinfo {volume} {111}},\ \bibinfo {pages} {056802} (\bibinfo {year}
  {2013})}\BibitemShut {NoStop}%
\bibitem [{\citenamefont {Douçot}\ and\ \citenamefont
  {Ioffe}(2007)}]{Ioffe2007}%
  \BibitemOpen
  \bibfield  {author} {\bibinfo {author} {\bibfnamefont {B.}~\bibnamefont
  {Douçot}}\ and\ \bibinfo {author} {\bibfnamefont {L.~B.}\ \bibnamefont
  {Ioffe}},\ }\href@noop {} {\bibfield  {journal} {\bibinfo  {journal} {Phys.
  Rev. B}\ }\textbf {\bibinfo {volume} {76}},\ \bibinfo {pages} {214507}
  (\bibinfo {year} {2007})}\BibitemShut {NoStop}%
\bibitem [{\citenamefont {Zazunov}\ \emph {et~al.}(2008)\citenamefont
  {Zazunov}, \citenamefont {Didier},\ and\ \citenamefont
  {Hekking}}]{Zazunov2008}%
  \BibitemOpen
  \bibfield  {author} {\bibinfo {author} {\bibfnamefont {A.}~\bibnamefont
  {Zazunov}}, \bibinfo {author} {\bibfnamefont {N.}~\bibnamefont {Didier}}, \
  and\ \bibinfo {author} {\bibfnamefont {F.~W.~J.}\ \bibnamefont {Hekking}},\
  }\href@noop {} {\bibfield  {journal} {\bibinfo  {journal} {EPL}\ }\textbf
  {\bibinfo {volume} {83}},\ \bibinfo {pages} {47012} (\bibinfo {year}
  {2008})}\BibitemShut {NoStop}%
\bibitem [{\citenamefont {Dartiailh}\ \emph {et~al.}(2020)\citenamefont
  {Dartiailh}, \citenamefont {Cuozzo}, \citenamefont {Mayer},\ and\
  \citenamefont {Yuan}}]{Shabani2020}%
  \BibitemOpen
  \bibfield  {author} {\bibinfo {author} {\bibfnamefont {M.~C.}\ \bibnamefont
  {Dartiailh}}, \bibinfo {author} {\bibfnamefont {J.~J.}\ \bibnamefont
  {Cuozzo}}, \bibinfo {author} {\bibfnamefont {W.}~\bibnamefont {Mayer}}, \
  and\ \bibinfo {author} {\bibfnamefont {J.}~\bibnamefont {Yuan}},\ }\href@noop
  {} {\bibfield  {journal} {\bibinfo  {journal} {arXiv:2005.00077}\ } (\bibinfo
  {year} {2020})}\BibitemShut {NoStop}%
\bibitem [{\citenamefont {Rodríguez-Mota}\ \emph {et~al.}(2019)\citenamefont
  {Rodríguez-Mota}, \citenamefont {Vishveshwara},\ and\ \citenamefont
  {Pereg-Barnea}}]{RodriguezMota2019}%
  \BibitemOpen
  \bibfield  {author} {\bibinfo {author} {\bibfnamefont {R.}~\bibnamefont
  {Rodríguez-Mota}}, \bibinfo {author} {\bibfnamefont {S.}~\bibnamefont
  {Vishveshwara}}, \ and\ \bibinfo {author} {\bibfnamefont {T.}~\bibnamefont
  {Pereg-Barnea}},\ }\href@noop {} {\bibfield  {journal} {\bibinfo  {journal}
  {Phys. Rev. B}\ }\textbf {\bibinfo {volume} {99}},\ \bibinfo {pages} {024517}
  (\bibinfo {year} {2019})}\BibitemShut {NoStop}%
\bibitem [{\citenamefont {Coleman}(1977)}]{Coleman1977}%
  \BibitemOpen
  \bibfield  {author} {\bibinfo {author} {\bibfnamefont {S.}~\bibnamefont
  {Coleman}},\ }in\ \href@noop {} {\emph {\bibinfo {booktitle} {Lectures
  delivered at the International School of Subnuclear Physics}}}\ (\bibinfo
  {publisher} {Erice},\ \bibinfo {year} {1977})\BibitemShut {NoStop}%
\bibitem [{\citenamefont {Matveev}\ \emph {et~al.}(2002)\citenamefont
  {Matveev}, \citenamefont {Larkin},\ and\ \citenamefont
  {Glazman}}]{Matveev2002}%
  \BibitemOpen
  \bibfield  {author} {\bibinfo {author} {\bibfnamefont {K.~A.}\ \bibnamefont
  {Matveev}}, \bibinfo {author} {\bibfnamefont {A.~I.}\ \bibnamefont {Larkin}},
  \ and\ \bibinfo {author} {\bibfnamefont {L.~I.}\ \bibnamefont {Glazman}},\
  }\href@noop {} {\bibfield  {journal} {\bibinfo  {journal} {Phys. Rev. Lett.}\
  }\textbf {\bibinfo {volume} {89}},\ \bibinfo {pages} {096802} (\bibinfo
  {year} {2002})}\BibitemShut {NoStop}%
\bibitem [{\citenamefont {Vainshtein}\ \emph {et~al.}(1982)\citenamefont
  {Vainshtein}, \citenamefont {Zakharov}, \citenamefont {Novikov},\ and\
  \citenamefont {Shifman}}]{Vainshtein1982}%
  \BibitemOpen
  \bibfield  {author} {\bibinfo {author} {\bibfnamefont {A.~I.}\ \bibnamefont
  {Vainshtein}}, \bibinfo {author} {\bibfnamefont {V.~I.}\ \bibnamefont
  {Zakharov}}, \bibinfo {author} {\bibfnamefont {V.~A.}\ \bibnamefont
  {Novikov}}, \ and\ \bibinfo {author} {\bibfnamefont {M.~A.}\ \bibnamefont
  {Shifman}},\ }\href@noop {} {\bibfield  {journal} {\bibinfo  {journal} {Sov.
  Phys. Uspekhi}\ }\textbf {\bibinfo {volume} {25}},\ \bibinfo {pages} {195}
  (\bibinfo {year} {1982})},\ \bibinfo {note} {[Usp. Fiz. Nauk. \textbf{136},
  553 (1982)]}\BibitemShut {NoStop}%
\bibitem [{\citenamefont {Dmytruk}\ and\ \citenamefont
  {Klinovaja}(2018)}]{Dmytruk2018}%
  \BibitemOpen
  \bibfield  {author} {\bibinfo {author} {\bibfnamefont {O.}~\bibnamefont
  {Dmytruk}}\ and\ \bibinfo {author} {\bibfnamefont {J.}~\bibnamefont
  {Klinovaja}},\ }\href@noop {} {\bibfield  {journal} {\bibinfo  {journal}
  {Phys. Rev. B}\ }\textbf {\bibinfo {volume} {97}},\ \bibinfo {pages} {155409}
  (\bibinfo {year} {2018})}\BibitemShut {NoStop}%
\bibitem [{\citenamefont {Hoffman}\ \emph {et~al.}(2017)\citenamefont
  {Hoffman}, \citenamefont {Chevallier}, \citenamefont {Loss},\ and\
  \citenamefont {Klinovaja}}]{Hoffman2017}%
  \BibitemOpen
  \bibfield  {author} {\bibinfo {author} {\bibfnamefont {S.}~\bibnamefont
  {Hoffman}}, \bibinfo {author} {\bibfnamefont {D.}~\bibnamefont {Chevallier}},
  \bibinfo {author} {\bibfnamefont {D.}~\bibnamefont {Loss}}, \ and\ \bibinfo
  {author} {\bibfnamefont {J.}~\bibnamefont {Klinovaja}},\ }\href@noop {}
  {\bibfield  {journal} {\bibinfo  {journal} {Phys. Rev. B}\ }\textbf {\bibinfo
  {volume} {96}},\ \bibinfo {pages} {045440} (\bibinfo {year}
  {2017})}\BibitemShut {NoStop}%
\bibitem [{\citenamefont {Rainis}\ and\ \citenamefont
  {Loss}(2012)}]{Rainis2012}%
  \BibitemOpen
  \bibfield  {author} {\bibinfo {author} {\bibfnamefont {D.}~\bibnamefont
  {Rainis}}\ and\ \bibinfo {author} {\bibfnamefont {D.}~\bibnamefont {Loss}},\
  }\href@noop {} {\bibfield  {journal} {\bibinfo  {journal} {Phys. Rev. B}\
  }\textbf {\bibinfo {volume} {85}},\ \bibinfo {pages} {174533} (\bibinfo
  {year} {2012})}\BibitemShut {NoStop}%
\bibitem [{\citenamefont {Schmidt}\ \emph {et~al.}(2012)\citenamefont
  {Schmidt}, \citenamefont {Rainis},\ and\ \citenamefont {Loss}}]{Schmidt2012}%
  \BibitemOpen
  \bibfield  {author} {\bibinfo {author} {\bibfnamefont {M.~J.}\ \bibnamefont
  {Schmidt}}, \bibinfo {author} {\bibfnamefont {D.}~\bibnamefont {Rainis}}, \
  and\ \bibinfo {author} {\bibfnamefont {D.}~\bibnamefont {Loss}},\ }\href@noop
  {} {\bibfield  {journal} {\bibinfo  {journal} {Phys. Rev. B}\ }\textbf
  {\bibinfo {volume} {86}},\ \bibinfo {pages} {085414} (\bibinfo {year}
  {2012})}\BibitemShut {NoStop}%
\bibitem [{\citenamefont {Higginbotham}\ \emph {et~al.}(2015)\citenamefont
  {Higginbotham}, \citenamefont {Albrecht}, \citenamefont {Kiršanskas},
  \citenamefont {Chang}, \citenamefont {Kuemmeth}, \citenamefont {Krogstrup},
  \citenamefont {Jespersen}, \citenamefont {Nygård}, \citenamefont
  {Flensberg},\ and\ \citenamefont {Marcus}}]{Higgenbotham2015}%
  \BibitemOpen
  \bibfield  {author} {\bibinfo {author} {\bibfnamefont {A.~P.}\ \bibnamefont
  {Higginbotham}}, \bibinfo {author} {\bibfnamefont {S.~M.}\ \bibnamefont
  {Albrecht}}, \bibinfo {author} {\bibfnamefont {G.}~\bibnamefont
  {Kiršanskas}}, \bibinfo {author} {\bibfnamefont {W.}~\bibnamefont {Chang}},
  \bibinfo {author} {\bibfnamefont {F.}~\bibnamefont {Kuemmeth}}, \bibinfo
  {author} {\bibfnamefont {P.}~\bibnamefont {Krogstrup}}, \bibinfo {author}
  {\bibfnamefont {T.~S.}\ \bibnamefont {Jespersen}}, \bibinfo {author}
  {\bibfnamefont {J.}~\bibnamefont {Nygård}}, \bibinfo {author} {\bibfnamefont
  {K.}~\bibnamefont {Flensberg}}, \ and\ \bibinfo {author} {\bibfnamefont
  {C.~M.}\ \bibnamefont {Marcus}},\ }\href@noop {} {\bibfield  {journal}
  {\bibinfo  {journal} {Nat. Phys.}\ }\textbf {\bibinfo {volume} {11}},\
  \bibinfo {pages} {1017} (\bibinfo {year} {2015})}\BibitemShut {NoStop}%
\bibitem [{\citenamefont {Albrecht}\ \emph {et~al.}(2017)\citenamefont
  {Albrecht}, \citenamefont {Hansen}, \citenamefont {Higginbotham},
  \citenamefont {Kuemmeth}, \citenamefont {Jespersen}, \citenamefont
  {Nyg\aa{}rd}, \citenamefont {Krogstrup}, \citenamefont {Danon}, \citenamefont
  {Flensberg},\ and\ \citenamefont {Marcus}}]{Albrecht2017}%
  \BibitemOpen
  \bibfield  {author} {\bibinfo {author} {\bibfnamefont {S.~M.}\ \bibnamefont
  {Albrecht}}, \bibinfo {author} {\bibfnamefont {E.~B.}\ \bibnamefont
  {Hansen}}, \bibinfo {author} {\bibfnamefont {A.~P.}\ \bibnamefont
  {Higginbotham}}, \bibinfo {author} {\bibfnamefont {F.}~\bibnamefont
  {Kuemmeth}}, \bibinfo {author} {\bibfnamefont {T.~S.}\ \bibnamefont
  {Jespersen}}, \bibinfo {author} {\bibfnamefont {J.}~\bibnamefont
  {Nyg\aa{}rd}}, \bibinfo {author} {\bibfnamefont {P.}~\bibnamefont
  {Krogstrup}}, \bibinfo {author} {\bibfnamefont {J.}~\bibnamefont {Danon}},
  \bibinfo {author} {\bibfnamefont {K.}~\bibnamefont {Flensberg}}, \ and\
  \bibinfo {author} {\bibfnamefont {C.~M.}\ \bibnamefont {Marcus}},\
  }\href@noop {} {\bibfield  {journal} {\bibinfo  {journal} {Phys. Rev. Lett.}\
  }\textbf {\bibinfo {volume} {118}},\ \bibinfo {pages} {137701} (\bibinfo
  {year} {2017})}\BibitemShut {NoStop}%
\bibitem [{\citenamefont {Aseev}\ \emph {et~al.}(2018)\citenamefont {Aseev},
  \citenamefont {Klinovaja},\ and\ \citenamefont {Loss}}]{Aseev2018}%
  \BibitemOpen
  \bibfield  {author} {\bibinfo {author} {\bibfnamefont {P.~P.}\ \bibnamefont
  {Aseev}}, \bibinfo {author} {\bibfnamefont {J.}~\bibnamefont {Klinovaja}}, \
  and\ \bibinfo {author} {\bibfnamefont {D.}~\bibnamefont {Loss}},\ }\href@noop
  {} {\bibfield  {journal} {\bibinfo  {journal} {Phys. Rev. B}\ }\textbf
  {\bibinfo {volume} {98}},\ \bibinfo {pages} {155414} (\bibinfo {year}
  {2018})}\BibitemShut {NoStop}%
\bibitem [{\citenamefont {Budich}\ \emph {et~al.}(2012)\citenamefont {Budich},
  \citenamefont {Walter},\ and\ \citenamefont {Trauzettel}}]{Budich2012}%
  \BibitemOpen
  \bibfield  {author} {\bibinfo {author} {\bibfnamefont {J.~C.}\ \bibnamefont
  {Budich}}, \bibinfo {author} {\bibfnamefont {S.}~\bibnamefont {Walter}}, \
  and\ \bibinfo {author} {\bibfnamefont {B.}~\bibnamefont {Trauzettel}},\
  }\href@noop {} {\bibfield  {journal} {\bibinfo  {journal} {Phys. Rev. B}\
  }\textbf {\bibinfo {volume} {85}},\ \bibinfo {pages} {121405(R)} (\bibinfo
  {year} {2012})}\BibitemShut {NoStop}%
\bibitem [{\citenamefont {Karzig}\ \emph {et~al.}(2008)\citenamefont {Karzig},
  \citenamefont {Cole},\ and\ \citenamefont {Pikulin}}]{Karzig2020}%
  \BibitemOpen
  \bibfield  {author} {\bibinfo {author} {\bibfnamefont {T.}~\bibnamefont
  {Karzig}}, \bibinfo {author} {\bibfnamefont {W.~S.}\ \bibnamefont {Cole}}, \
  and\ \bibinfo {author} {\bibfnamefont {D.~I.}\ \bibnamefont {Pikulin}},\
  }\href@noop {} {\bibfield  {journal} {\bibinfo  {journal} {arXiv:2004.01264}\
  } (\bibinfo {year} {2008})}\BibitemShut {NoStop}%
\bibitem [{\citenamefont {Pikulin}\ and\ \citenamefont
  {Nazarov}(2012)}]{Pikulin2012}%
  \BibitemOpen
  \bibfield  {author} {\bibinfo {author} {\bibfnamefont {D.~I.}\ \bibnamefont
  {Pikulin}}\ and\ \bibinfo {author} {\bibfnamefont {Y.~V.}\ \bibnamefont
  {Nazarov}},\ }\href@noop {} {\bibfield  {journal} {\bibinfo  {journal} {Phys.
  Rev. B.}\ }\textbf {\bibinfo {volume} {86}},\ \bibinfo {pages} {140504(R)}
  (\bibinfo {year} {2012})}\BibitemShut {NoStop}%
\bibitem [{\citenamefont {Dom{\'i}nguez}\ \emph {et~al.}(2012)\citenamefont
  {Dom{\'i}nguez}, \citenamefont {Hassler},\ and\ \citenamefont
  {Platero}}]{Dominguez2012}%
  \BibitemOpen
  \bibfield  {author} {\bibinfo {author} {\bibfnamefont {F.}~\bibnamefont
  {Dom{\'i}nguez}}, \bibinfo {author} {\bibfnamefont {F.}~\bibnamefont
  {Hassler}}, \ and\ \bibinfo {author} {\bibfnamefont {G.}~\bibnamefont
  {Platero}},\ }\href@noop {} {\bibfield  {journal} {\bibinfo  {journal} {Phys.
  Rev. B.}\ }\textbf {\bibinfo {volume} {86}},\ \bibinfo {pages} {140503(R)}
  (\bibinfo {year} {2012})}\BibitemShut {NoStop}%
\bibitem [{\citenamefont {Feng}\ \emph {et~al.}(2018)\citenamefont {Feng},
  \citenamefont {Huang}, \citenamefont {Wang},\ and\ \citenamefont
  {Niu}}]{Feng2018}%
  \BibitemOpen
  \bibfield  {author} {\bibinfo {author} {\bibfnamefont {J.-J.}\ \bibnamefont
  {Feng}}, \bibinfo {author} {\bibfnamefont {Z.}~\bibnamefont {Huang}},
  \bibinfo {author} {\bibfnamefont {Z.}~\bibnamefont {Wang}}, \ and\ \bibinfo
  {author} {\bibfnamefont {Q.}~\bibnamefont {Niu}},\ }\href@noop {} {\bibfield
  {journal} {\bibinfo  {journal} {Phys. Rev. B.}\ }\textbf {\bibinfo {volume}
  {98}},\ \bibinfo {pages} {134515} (\bibinfo {year} {2018})}\BibitemShut
  {NoStop}%
\bibitem [{\citenamefont {San-Jose}\ \emph {et~al.}(2012)\citenamefont
  {San-Jose}, \citenamefont {Prada},\ and\ \citenamefont
  {Aguado}}]{San-Jose2012}%
  \BibitemOpen
  \bibfield  {author} {\bibinfo {author} {\bibfnamefont {P.}~\bibnamefont
  {San-Jose}}, \bibinfo {author} {\bibfnamefont {E.}~\bibnamefont {Prada}}, \
  and\ \bibinfo {author} {\bibfnamefont {R.}~\bibnamefont {Aguado}},\
  }\href@noop {} {\bibfield  {journal} {\bibinfo  {journal} {Phys. Rev. Lett.}\
  }\textbf {\bibinfo {volume} {108}},\ \bibinfo {pages} {257001} (\bibinfo
  {year} {2012})}\BibitemShut {NoStop}%
\bibitem [{\citenamefont {Rainis}\ \emph {et~al.}(2013)\citenamefont {Rainis},
  \citenamefont {Trifunovic}, \citenamefont {Klinovaja},\ and\ \citenamefont
  {Loss}}]{Rainis2013}%
  \BibitemOpen
  \bibfield  {author} {\bibinfo {author} {\bibfnamefont {D.}~\bibnamefont
  {Rainis}}, \bibinfo {author} {\bibfnamefont {L.}~\bibnamefont {Trifunovic}},
  \bibinfo {author} {\bibfnamefont {J.}~\bibnamefont {Klinovaja}}, \ and\
  \bibinfo {author} {\bibfnamefont {D.}~\bibnamefont {Loss}},\ }\href@noop {}
  {\bibfield  {journal} {\bibinfo  {journal} {Phys. Rev. B}\ }\textbf {\bibinfo
  {volume} {87}},\ \bibinfo {pages} {024515} (\bibinfo {year}
  {2013})}\BibitemShut {NoStop}%
\end{thebibliography}%
\end{document}